\documentclass[twocolumn]{aastex631}
\usepackage{color}
\usepackage{hyperref}
\usepackage{natbib}

\begin{document}

% TeX/LateX macro definitions for solar physics book
%
% Last modified: 1999 April 17, PC
%==========================================================
%\textwidth=7.0truein
%\textheight=8.75truein
%---------- some extra title macros

\def\appendice#1#2{\vfill\eject{\Large \centerline{\bf APPENDICE {#1}.}
   \bigskip\noindent{\centerline{\bf {#2}}}}\bigskip\hrule\bigskip}
\def\annotbib{\bigskip\hrule\bigskip\noindent{\bf Bibliographie:}\bigskip}
\def\problems{\bigskip\hrule\bigskip\noindent{\bf Exercices:}\bigskip}
\def\lectures{\bigskip\hrule\bigskip\noindent{\bf Lectures suppl\'ementaires:}\bigskip}

%---------- this is a temporary "box" macros, until I figure out
%           how to set up a real box
% \def\boxdir{/home/paulchar/course/book/text/boxes}
\def\partie#1#2{
   \vfill\eject $~$\vskip 1.0truein{\centerline{\Huge\bf {#1}}
   \bigskip\bigskip\bigskip
   {\centerline{\Huge\bf {#2}}}}\vskip 1.5truein}
\def\boxdir{boxes}
\def\ourbox#1#2#3{\medskip\noindent\medskip\noindent\hrule\medskip
\noindent{\bf Box {#1}: {#2}}
\hfill\break\noindent{\input {\boxdir/#3}}
\medskip\hrule\bigskip
}

%---------- derivative macros

\def\derp#1#2{ { \partial #1  \over\partial #2 } } % dA/dB-type derivatives
\def\dtrp#1#2{ { \partial^2 #1\over\partial {#2}^2 } }
\def\dert#1#2{ { {\rm d}#1    \over {\rm d}#2 } }
\def\dtrt#1#2{ { {\rm d}^2 #1 \over{\rm d}{#2}^2 } }
\def\Dert#1#2{ { {\rm D}#1    \over {\rm D}#2 } }

\def\tderp#1#2{ \partial #1  /\partial #2  }   % "Text" dA/dB derivatives
\def\tdtrp#1#2{ \partial^2 #1/\partial {#2}^2}
\def\tdert#1#2{ {\rm d}#1    / {\rm d}#2 }
\def\tdtrt#1#2{ {\rm d}^2 #1 /{\rm d}{#2}^2  }
\def\tDert#1#2{ {\rm D}#1    / {\rm D}#2  }

\def\ddp#1{{\partial\over\partial#1}}          % d/dA-type derivatives
\def\ddt#1{{{\rm d}\over {\rm d}#1}}
\def\DDt#1{{{\rm D}\over {\rm D}#1}}
\def\ddtp#1{{\partial^2\over\partial#1^2}}
\def\ddtt#1{{{\rm d}^2\over {\rm d}#1^2}}
\def\tddp#1{\partial/\partial#1} 
\def\tddt#1{{\rm d}/{\rm d}#1} 
\def\tDDt#1{{\rm D}/{\rm D}#1} 
\def\rd{{\rm d}}                               % roman-font d, for math mode
\def\dd#1{\,{\rm d}{#1}}                         % roman-font d, for math mode

%---------- mathematical operators ans symbols

\def\vec#1{{\bf #1}}                               % vector
\def\snorm#1{\mid #1\mid}                          % scalar norm
\def\vnorm#1{\|#1\|}                               % vector norm
\def\norm{\hat {\bf n}}                            % vector norm
\def\uvec#1{\hat {\bf e}_{#1}}                       % unit vector
\def\avrg#1{\left<{#1}\right>}                     % averaging brackets
\def\max{_{\rm max}}                               % max subscript
\def\min{_{\rm min}}                               % min subscript
\def\dS{{\rm d}S}                                  % surface element
\def\dV{{\rm d}V}                                  % volume element

%---------- physical units

\def\gcmds{{\rm g}\,{\rm cm}^{-2}\,{\rm s}^{-1}}   % mass flux
\def\ecmds{{\rm erg}\,{\rm cm}^{-2}\,{\rm s}^{-1}} % energy flux
\def\gcm3{{\rm g}\,{\rm cm}^{-3}}                  % mass density
\def\ecm3{{\rm erg}\,{\rm cm}^{-3}}                % energy density
\def\kms{{\rm km}\,{\rm s}^{-1}}                   % speed
\def\cms{{\rm cm}\,{\rm s}^{-1}}                   % speed
\def\rads{{\rm rad}\,{\rm s}^{-1}}                 % angular velocity
\def\cmds{{\rm cm}^2\,{\rm s}^{-1}}                % transport coeffs.
\def\cmms{{\rm cm}^{-2}\,{\rm s}^{-1}}             % particle flux
\def\cmt{{\rm cm}^{-3}}                            % unit volume

%---------- physical symbols

\def\Rs{R}                               % solar radius
\def\Ms{M}                               % solar mass
\def\Ls{L}                               % solar luminosity
\def\me{m_e}                             % electron mass
\def\mH{m_H}                             % Hydrogen mass
\def\rhoe{\rho_e}                        % Charge density
\def\Vsini{v\,{\rm sin}\,i}              % v.sin(i)
\def\gpot{\Phi}                          % Gravitational Potential
\def\sv#1{u_{#1}}                        % scalar velocity component
\def\vv{{\bf u}}                         % vector velocity
\def\Rm{{\rm R}_m}                       % Magnetic Reynolds number
\def\Rnu{{\rm R}_\nu}                    % Viscous Reynolds number
\def\Dcrit{D_{\rm crit}}                 % Critical dynamo number
\def\Beq{B_{\rm eq}}                     % Equipartition field strength
\def\Emag{{\cal E}_{\rm B}}              % Magnetic Energy
\def\Hmag{{\cal H}_{\rm B}}              % Magnetic Helicity
\def\turnov {\tau_{\rm c}}               % Turnover time
\def\tdmag {\tau_{\eta}}                 % Magnetic dissipation time

%---------- mathematical symbols

\def\Real{\leavevmode\hbox{I\kern-.2115em\hbox{R}}}        % Real numbers
\def\spose#1{\hbox to 0pt{#1\hss}}
\def\lta{\mathrel{\spose{\lower 3pt\hbox{$\mathchar"218$}} % less-than-about
     \raise 2.0pt\hbox{$\mathchar"13C$}}}
\def\gta{\mathrel{\spose{\lower 3pt\hbox{$\mathchar"218$}} % greater-than-about
     \raise 2.0pt\hbox{$\mathchar"13E$}}}

%---------- this is for references in bibliography

\def\ourref#1{\hangindent=6em\hangafter=1 {#1}}
\def\etal{{\it et al.}}
\def\AandA#1#2{{\it Astron.~Ap.}, {\bf #1}, #2}
\def\ApJ#1#2{{\it Astrophys.~J.}, {\bf #1}, #2}
\def\ApN#1#2{{\it Astrophys.~Nach.}, {\bf #1}, #2}
\def\ARAA#1#2{{\it Ann.~Rev.~Astron.~Ap.}, {\bf #1}, #2}
\def\ARFM#1#2{{\it Ann.~Rev.~Fluid Mech.}, {\bf #1}, #2}
\def\CMP#1#2{{\it Comm.~Math.~Phys.}, {\bf #1}, #2}
\def\GAFD#1#2{{\it Geophys.~Astrophys.~Fluid Dyn.}, {\bf #1}, #2}
\def\GRL#1#2{{\it Geophys.~Res.~Lett.}, {\bf #1}, #2}
\def\JCompP#1#2{{\it J.~Comp.~Phys.}, {\bf #1}, #2}
\def\JFM#1#2{{\it J.~Fluid Mech.}, {\bf #1}, #2}
\def\JGR#1#2{{\it J.~Geophys.~Res.}, {\bf #1}, #2}
\def\MNRAS#1#2{{\it Mon.~Not.~Roy.~Astron.~Soc.}, {\bf #1}, #2}
\def\Nature#1#2{{\it Nature}, {\bf #1}, #2}
\def\PR#1#2{{\it Phys.~Rev.}, {\bf #1}, #2}
\def\PRA#1#2{{\it Phys.~Rev.~A}, {\bf #1}, #2}
\def\PRE#1#2{{\it Phys.~Rev.~E}, {\bf #1}, #2}
\def\PRL#1#2{{\it Phys.~Rev.~Lett.}, {\bf #1}, #2}
\def\RevMP#1#2{{\it Rev.~Mod.~Phys.}, {\bf #1}, #2}
\def\Science#1#2{{\it Science}, {\bf #1}, #2}
\def\SolP#1#2{{\it Solar Phys.}, {\bf #1}, #2}
\def\ZAp#1#2{{\it Zeitschrift Ap.}, {\bf #1}, #2}

%---------- this is Tom 's trick for boldface greek letters and such

\def\bfce{\setbox2=\hbox{$\ce$}
\hbox{{$\ce$}\hskip-.97\wd2
{$\ce$}\hskip-.97\wd2
{$\ce$}\hskip-.97\wd2 {$\ce$}}}

% ceci pour Omega-vecteur
\def\bfOmega{\setbox2=\hbox{$\Omega$}
\hbox{{$\Omega$}\hskip-.97\wd2
{$\Omega$}\hskip-.97\wd2
{$\Omega$}\hskip-.97\wd2 {$\Omega$}}}

% ceci pour omega-vecteur
\def\bfomega{\setbox2=\hbox{$\omega$}
\hbox{{$\omega$}\hskip-.97\wd2
{$\omega$}\hskip-.97\wd2
{$\omega$}\hskip-.97\wd2 {$\omega$}}}

% ceci pour xi-vecteur
\def\bfxi{\setbox2=\hbox{$\xi$}
\hbox{{$\xi$}\hskip-.97\wd2
{$\xi$}\hskip-.97\wd2
{$\xi$}\hskip-.97\wd2 {$\xi$}}}

% ceci pour Xi-vecteur
\def\bfXi{\setbox2=\hbox{$\Xi$}
\hbox{{$\Xi$}\hskip-.97\wd2
{$\Xi$}\hskip-.97\wd2
{$\Xi$}\hskip-.97\wd2 {$\Xi$}}}

% ceci pour beta-vecteur
\def\bfbeta{\setbox2=\hbox{$\beta$}
\hbox{{$\beta$}\hskip-.97\wd2
{$\beta$}\hskip-.97\wd2
{$\beta$}\hskip-.97\wd2 {$\beta$}}}

% ceci pour tau-vecteur
\def\bftau{\setbox2=\hbox{$\tau$}
\hbox{{$\tau$}\hskip-.97\wd2
{$\tau$}\hskip-.97\wd2
{$\tau$}\hskip-.97\wd2 {$\tau$}}}

% ceci pour ell-vecteur
\def\bfell{\setbox2=\hbox{$\ell$}
\hbox{{$\ell$}\hskip-.97\wd2
{$\ell$}\hskip-.97\wd2
{$\ell$}\hskip-.97\wd2 {$\ell$}}}

%======================================================================

\def\pref#1{\protect{\ref{#1}}}

\title{Breaking gyrochronology through the collapse of coronal winds}

\author[0009-0001-2367-2243]{Micha\"el L\'evesque}
\affiliation{D\'epartement de Physique, Universit\'e de Montr\'eal, Montr\'eal, QC, H3C 3J7, Canada\\}
\email{michael.levesque.1@umontreal.ca}
\author[0000-0003-1618-3924]{Paul Charbonneau}
\affiliation{D\'epartement de Physique, Universit\'e de Montr\'eal, Montr\'eal, QC, H3C 3J7, Canada\\}
\email{paul.charbonneau@umontreal.ca}

%\maketitle
\begin{abstract}
Gyrochronology, a method for dating aged field stars ($\gta$ a few Gyr) based on their rotation rate, has recently been shown to fail for many stars older than the sun. The explanation most often put forth is that a shutdown or mode change in the stellar dynamo leads to a sharp decrease in angular momentum loss in magnetized coronal winds. In this paper, we explore an alternate possibility, namely a collapse of
the wind itself through a reduction of coronal heating. We show that in the low coronal temperature ($T_0$) limit, even at solar-like
low rotation rates ($\Omega$) and coronal magnetic field strength ($B_{r0}$), magnetocentrifugal effects are important and preclude expression of the mass and angular momentum loss rates as power-laws of $T_0$ or $\Omega$ when $T_0$ drops below $\simeq 1.5\,$MK. Mass loss is found to scale linearly with power input into the wind at all coronal temperatures. Introducing an ad hoc power law relationship $T_0\propto B_{r0}^\sigma$ while retaining the ``standard'' dynamo relationship $B_{r0}\propto\Omega$, we show that reproducing the observed break in gyrochronology requires an exponent
$\sigma\gta 1.5$, with which is associated a drop by over 3 orders of magnitude in power input into the quiet corona. This appears physically unrealistic, given current observations of chromospheric and coronal non-thermal emission in aged solar-type stars.
\end{abstract}

\section{Introduction}

In a short and epoch-making paper, \cite{Skumanich1972} gathered observations of
rotation rates in main-sequence solar-type stars of known ages, and argued that their
rotation rate decreases as the inverse square root of their age, a relationship since
known as Skumanich's Law. He also showed that emission in the core of the calcium H and K lines, a know proxy of surface magnetism, also decreased with stellar
age following a similar time dependence. This remarkable $t^{-1/2}$ relationship has been harnessed to establish the field of gyrochronology \citep[see, e.g.][and reference
therein]{Barnes2003,Isiketal2023}, namely using rotation rate as a proxy
for stellar age.

The Skumanich Law has withstood the test of time, and is now considered to
hold from ages of $\sim 1$--$2\,$Gyr to the solar age ($\simeq 4.5\,$Gyr), with the lower
bound depending on stellar mass to some extent \citep[][and references therein]{GalletBouvier2015,LanzafameSpada2015}. It therefore came as a surprise
when a number of solar-type field stars slightly older than the sun, when dated through asteroseismology, were found to rotate significantly faster than predicted by Skumanich's Law
\citep{vanSadersetal2016,Metcalfeetal2023,Saundersetal2024}, suggesting a rapid decrease of the angular momentum loss rate (${\dot J})$ beyond the solar age, as compared to the expectation based on the extrapolation of Skumanich's Law.

One possible explanation for this sudden drop in ${\dot J}$, and that emphasized
thus far \citep[e.g.,][and references therein]{Metcalfeetal2023}, is a shutdown or mode change in
the large-scale dynamo producing the global coronal magnetic field responsible for enhancing angular momentum loss over that characterizing unmagnetized coronal winds.
%(more on this in \S\ref{ssec:WDwind} below). 
This would imply that the current solar dynamo 
is running very close to criticality
%, or some other bifurcation 
\citep[e.g.,][]{Metcalfeetal2016}.
%, an idea finding supports from a variety of arguments; \textcolor{blue}{global hydrodynamical and magnetohydrodynamical simulations of solar convection suggests that the latter operates in a Rossby number regime close to the transition between ``solar-like'' differentialrotation, characterized by equatorial regions rotating more rapidly than the poles, and the so-called ``anti-solar'' regime, in which equatorial regions rotate more slowly than mid-latitudes \citep[][and references therein]{Guerreroetal2013,Gastineetal2014,Mabuchietal2015,HottaKusano2021}. Differential rotation being generally considered as an important ingredient (and energy source) for the solar dynamo \citep{Charbonneau2020},  transiting from solar-like to anti-solar differential rotation would presumably involve, at some point, going through a phase of much reduced differential rotation, with concomittant reduction of the dynamo number possibly below its critical value.} \textcolor{red}{The bits in blue could go, replaced by 1--2 REFs discussing this}
An alternate, related explanation not involving a full shutdown of the large-scale dynamo could be 
a bifurcation-like switch of its mode of operation, leading to a dominance 
of higher-order multipoles over the dipolar mode, with a potential large impact
on the angular momentum loss rate
\citep[][and references therein]{FinleyMatt2018,Garraffoetal2018,Metcalfeetal2023}.

In this paper we explore an alternate means of shutting off the angular momentum loss rate, namely a collapse of coronal wind triggered by reduced energy input
to coronal heating. The idea that a drop in coronal temperature can lower
angular momentum loss in coronal wind through a strong reduction 
of the associated mass loss rate has already been explored using various empirical and semi-empirical models 
\citep[see, e.g.,][]{PantolmosMatt2017,OFionnagainVidotto2018}, 
which have demonstrated the potential viability of this explanatory framework.
%\textcolor{blue}{Our approach, however, differs from these earlier works: we use a MHD wind model which, although geometrically simplified and restricted to the simplest reasonable magnetic field topology, is computationally efficient enough to allow wide exploration of parameter space, while capturing without approximations or simplifications the magnetocentrifugal contributions to wind driving. Perhaps more importantly, we quantify the hypothesis of coronal wind collapse in terms of coronal heating rates. We also examine the problem in the context of a bona fide spin-down model, so that the rotational history of the star is also taken into consideration.}
%\textcolor{red}{Preceeding bit in blue could be omitted}
%\smallskip
%Not clear if "break" is gradual or near-discontinuous. Saunders et al 2024
%\smallskip
%Break happens at different $\Omega$ for different stellar masses ? Look up REFS ?

The remainder of the paper is organized as follows: we begin
(\S\ref{sec:SkuLaw}) by recalling in some detail the derivation of Skumanich's Law from wind theory, with emphasis on the various physical assumptions required to arrive at $\Omega\propto t^{-1/2}$. We turn in \S\ref{sec:lowT} to the dynamics of winds at coronal base temperatures approaching the hydrostatic corona limit, and on this basis we examine in \S\ref{sec:breakSku} under which conditions Skumanich's law can be broken in solar-type stars slightly older than the sun. 
  In \S\ref{sec:Wenergetics} we revisit the results of \S\ref{sec:breakSku} from the point of view of wind energetics and coronal heating, seeking to determine whether scenarios succeeding in breaking Skumanich's Law are energetically consistent with current ideas and observational constraints. We conclude in \S\ref{sec:conclu} by summarizing our main results, and placing them in the context of alternate, dynamo-based explanations for the break of gyrochronology in middle-aged solar-type stars. 

\section{Magnetized coronal winds and the Skumanich $t^{-1/2}$ Law\label{sec:SkuLaw}}

Magnetized wind outflows emanating from the coronae of solar type stars carry away angular
momentum, and thus exert a torque $\Gamma$ that slows the star's rotation
\citep{Schatzman1962,WeberDavis1967,Mestel1968}.
Assume for now that the stellar moment of inertia ($I$) remains constant
on the main-sequence and that the star rotates
as a solid-body at angular velocity $\Omega$; the stellar angular momentum 
content is then simply $J=I\Omega$, with ${\dot J}=-\Gamma$ (here and in what follows, 
the dot indicates time derivative).
The specific case
$\Gamma\propto\Omega^3$ \citep[see][and \S 2.2 below]{Durney1972} then yields ${\dot\Omega}\propto -\Omega^3$,
which integrates to
\begin{equation}
{1\over\Omega^2(t)}-{1\over\Omega_0^2}\propto t-t_0~,
\end{equation}
with initial condition $\Omega_0\equiv\Omega(t_0)$
on the ZAMS ($t_0$). 
In the late phases of spindown, i.e., $t\gg t_0$ and $\Omega(t)\ll\Omega_0$,
Skumanich's Law is recovered:
\begin{equation}
\Omega(t)\propto t^{-1/2}~.
\end{equation}
The missing
constants of proportionality in the above expressions, including the
$\Gamma\propto\Omega^3$ Ansatz, 
is of course where physics is hiding; they are determined by
stellar structure, setting the stellar moment of inertia, by
dynamo theory, which yields the global coronal magnetic field strength
and multipole configuration, and
by coronal wind speeds and associated mass loss, ultimately set by coronal heating.

Skumanich's Law was established observationally, but it can also be
derived from coronal wind theory under a set of specific physical assumptions.
As first shown by \cite{Durney1972},
three distinct assumptions lead to Skumanich's Law:

\begin{enumerate}
\item A ``dynamo relationship'' relating the base coronal magnetic field $B_0$
linearly to the rotation rate $\Omega$;
\item A coronal wind that is primarily thermally-driven;
\item A constant stellar moment of inertia (requiring solid-body rotation on the main-sequence).
\end{enumerate}

In the remainder of this section we
examine critically the physical underpinnings of these three assumptions.

\subsection{The dynamo relationship\label{ssec:dynamo}}

The ``dynamo relationship'' relating a star's global large-scale surface
magnetic field strength to its rotation rate is a crucial input ingredient for any rotational evolution model for solar-type stars. The sun offers an obvious anchor point to establish such a relationship, 
but even there one runs into major difficulties: at this writing there
still exists no consensus model for the solar dynamo, although many acceptable
models can be found in the literature \citep{Charbonneau2020}. 
While differential rotation is usually considered a key ingredient for the generation of the toroidal component of the
sun's large-scale magnetic field, no such consensus exists with regards to the
inductive mechanism amplifying (and reversing) the sun's large-scale dipole
\citep{CharbonneauSokoloff2023}; nor
regarding the specific workings of nonlinear backreaction
of the magnetic field on the internal flows, responsible for stabilizing the solar cycle amplitude to some more or less stable mean value, or for the mechanism(s) ---stochastic or deterministic--- responsible for producing the observed cycle-to-cycle amplitude variations
\citep{Charbonneau2020,Karak2023}.

Even if a consensus did exist on a ``solar dynamo model'', 
extending such a model from the sun 
to younger or older solar-type stars also faces a number of conceptual
difficulties. Even if stellar masses ---and thus overall
internal structure--- remain
close to solar, one must specify how large-scale flows such as 
differential rotation and meridional circulation 
(the latter also playing an important role in many contemporary solar dynamo models) 
vary with rotation rate and convective luminosity; 
also how turbulent induction and diffusion, set by the properties of turbulent convection, vary with rotation and luminosity. 
Theory and numerical simulations can (and do)
provide useful guidance 
\citep[see, e.g.][and references therein]{KitchatinovRuediger1999,Kuekeretal2011,Simardetal2016,Varelaetal2016,Warnecke2018,Brunetal2022}, but there remains still
far too many free parameters and unknowns to allow a physically sound
extrapolation from the sun.

Observationally, there is ample indication that stellar magnetic activity
increases with increasing rotation rate. \cite{Skumanich1972} already noted
that emission in the core of the H and K lines of Calcium, a know proxy
of magnetic activity on the sun, also decreased as
$t^{-1/2}$, on par with rotation \citep[see also][]{Brownetal2022}. Observations of stellar coronal X-Ray emission,
another proxy of surface magnetism, as well as direct measurements of average surface magnetic fields,
are also roughly consistent with a near-linear
increase of surface magnetic activity with rotation, holding in slow rotators
($\Omega/\Omega_\odot\lta 10$) but flattening out at higher rotation rates
\citep[][and references therein]{Reinersetal2022,Isiketal2023}. In such slow rotators, the data is consistent with a power-law
relationship $\langle B_\ast \rangle\propto {\rm Ro}^{-1.25}$, where the Rossby
number ${\rm Ro}\propto \Omega^{-1}$ at fixed convective turnover time.
%\textcolor{red}{[CHECK ALSO JEFFERS, discuss Brownetal2022]}

With large-scale flows and turbulent induction ultimately driven by the 
rotational influences on turbulent convection, a dependence of the
dynamo-generated global magnetic field strength on rotation is certainly expected; 
yet the causal chain between the dynamo and CaHK or X-Ray emission is long 
and complex. 
In particular, on the sun non-thermal emission is dominated by active regions 
and smaller photospheric magnetic structures, while what matters
for angular momentum loss in coronal winds is the strength of the
lowest order magnetic multipoles, which determine the structure 
of the coronal magnetic field on scales of order of the solar radius and larger
\citep[][and references therein]{Revilleetal2015,Seeetal2019}.

With all these caveat firmly in mind, in everything that follows
we retain the (common) dynamo relationship $B\propto\Omega$ 
as a zeroth-order approximation for middle-age slowly rotating solar-type stars, and proceed.

\subsection{The Weber-Davis MHD wind model\label{ssec:WDwind}}

%Hypothesis \#2: thermally-driven magnetized wind (WD model)
We adopt in what follow the magnetohydrodynamical coronal wind model
of \cite{WeberDavis1967} (hereafter WD; see also \citealt{BelcherMacGregor1976}
and \citealt{Sakurai1985}). 
Formulated in
spherical polar coordinates $(r,\theta,\phi)$, this is defined as 
an axisymmetric ($\partial/\partial\phi=0$) equatorial plane 
($\theta=\pi/2$) steady-state ($\partial/\partial t=0$) ideal MHD wind model, 
%with variable dependencies:
%
%\begin{eqnarray}
%{\bf B}&=&(B_r(r),0,B_\phi(r))~,\nonumber\\ {\bf u}&=&(u_r(r),0,u_\phi(r))~,\nonumber \\
%p&=& p(r)~,\nonumber\\
%\rho&=&\rho(r)~,
%\end{eqnarray}
%
in $GM_\odot/r^2$ gravity and with an imposed (split-)monopolar radial field, $B_r(r)\propto 1/r^2$.
The latter is actually a fairly realistic
representation of the observed solar coronal magnetic field at activity minimum, where the 
axisymmetric belt of coronal loops straddling the equator
is pulled open radially by the outflowing solar wind a few solar radii 
above the photosphere \citep[see, e.g.,][and references therein]{Revilleetal2015}.
Energy input is effectively
achieved by imposing a polytropic relationship $p\propto\rho^\alpha$,
with $1\leq \alpha\leq 5/3$ for an ideal gas.
As detailed in \S\ref{sec:Wenergetics} below, the polytropic assumption amounts to a specific profile of volumetric heating through the corona and wind; in the absence of a consensus
model for coronal heating, the polytropic approximation is in fact commonly used in the majority of
extant MHD coronal wind models formulated in two and three spatial dimensions 
\citep[see, e.g.][]{WashimiShibata1993,KeppensGoedbloed1999,Vidottoetal2009,Revilleetal2015,FinleyMatt2018}.
The governing equations are solved above a reference radius $r_0$ 
($=1.15R_\odot$ in all that follows), taken to correspond to the heliocentric radius of peak temperature in the low corona, where density $\rho_0$ and temperature $T_0$ are specified.

Its geometrical simplification notwithstanding, the WD model captures in a dynamically consistent manner the magnetocentrifugal driving of the winds, 
as well as the torque it exerts on the photosphere. 
Its predictions are well-validated against in situ interplanetary wind measurements at Earth's orbit \citep{Pizzoetal1983}, and compare well to numerical simulations using more realistic magnetic field configurations \citep[e.g.,][]{Revilleetal2015}. In such 2D or 3D models, as long as coronal arcades are opened by the solar wind at heliocentric radii much smaller than the Alfv\'en radius, the torque density remains close to the prediction of the WD model. However, the mass loss rate is reduced in proportion to the fraction of the solar surface threaded by magnetic fieldlines closing back onto the photosphere \citep[see][and references therein]{PantolmosMatt2017,Ahuiretal2020}.

The steady-state wind solutions are characterized by three invariants, which define conserved quantities along streamlines: the
mass flux $r^2\rho u_r$ from the continuity equation, the magnetic flux $r^2B_r$
from the $r$-component of the induction equation,
and a third invariant, $r(u_rB_\phi-u_\phi B_r)$
arising from the $\phi$-component of induction equation.
These invariants allow to integrate the
$\phi$-component of the momentum equation to yield a fourth invariant:
\begin{equation}
\label{eq:amloss0}
L=ru_\phi-{rB_\phi B_r\over \mu_0\rho u_r}
\end{equation}
expressing conservation of angular momentum carried away by the wind, as comprised
of two contributions: the specific angular momentum of the ouflowing plasma
(first term on RHS), and the magnetic torque density (second term on RHS).

Introducing the Alfv\'en speed ${\bf A}={\bf B}/\sqrt{\mu_0\rho}$, and
making good use of (\ref{eq:amloss0}), the aforementioned third
invariant can be recast in the form:
\begin{equation}
\label{eq:amlossrA}
u_\phi=\Omega r{(u_r^2L/\Omega r^2)-A_r^2\over u_r^2-A_r^2}
\end{equation}
which diverges at the Alfv\'en radius $r_A$ (where $u_r=A_r$), 
unless the numerator also vanishes, which then requires:
\begin{equation}
\label{eq:amloss1}
L=\Omega r_A^2
\end{equation}
\citep{WeberDavis1967,Mestel1968}.
This is a truly remarkable result: it is \emph{as if} magnetic stresses
force the outflowing plasma to co-rotate with the photosphere out to the Alfv\'en radius
$r_A$, after which it is released to a Keplerian orbit. With $r_A\gg r_0$
even in a slowly rotating and (relatively) weakly magnetized star as the sun,
the loss of angular momentum is vastly enhanced, over the specific angular momentum that would be extracted by an unmagnetized wind of the same mass loss rate.

We follow in this paper the solution procedure described in \cite{BelcherMacGregor1976}; see see also
\cite{Charbonneau1995}, \S 4, as well as
Appendix B herein, for details.
%In analogy to Eq.~(\ref{eq:amlossrA}),
%the $r$-component of the momentum equation is recast in an expression evincing
%two critical points, where the wind speed becomes equal to the phase speed
%of the slow and fast magnetosonic wave modes
%\citep[see Eq.~(6) in][]{BelcherMacGregor1976}.
%The wind solution is obtained
%by requiring it to pass smoothly through both these critical points, while
%conserving total volumetric energy. This requirement yields
%a system of 6 nonlinear coupled algebraic equations
%for the base flow speed (radial and zonal components), and radii of 
%and flow speeds at the slow and fast magnetosonic points, 
%from which the full wind solution can be reconstructed
%. A gradient-based scheme such as the Newton-Raphson method can be used for solving this strongly nonlinear multivariate root fiding problem, but the
%small radius of convergence in most of parameter space renders the numerical
%solution quite delicate 
%\citep[see][also \citealt{charbonneau1995}, \S 4, and Appendix below]{BelcherMacGregor1976}.

For solar parameters, magnetocentrifugal driving of the wind is weak,
contributing only a few percent of the thermal pressure force propelling 
the wind; however, magnetic torques still completely dominates ${\dot J}$,
via the second term on the RHS of
eq.~(\ref{eq:amloss0}). The time rate of change of angular momentum is
given by
\begin{eqnarray}
{\dot J} = {2\over 3}{\dot M}\times L
= -{8\pi\over 3}\rho_Au_{rA}\Omega r_A^4~,
\label{eq:amloss2}
\end{eqnarray}
where the
factor $2/3$ results from the projection and integration
of the equatorial WD solution over spherical shells, and
the second equality from (\ref{eq:amloss1}) and evaluating the mass flux 
at the Alfv\'en radius $r_A$; at that location, 
$u_{rA}=A_{rA}=B_{rA}/\sqrt{\mu_0\rho_A}$;
conservation of magnetic flux also imposes $r_0^2B_{r0}=r_A^2B_{rA}$, 
so that eq.~(\ref{eq:amloss2}) can be rewritten as
\begin{equation}
{\dot J} = -{8\pi\over 3\mu_0}r_0^4B_{r0}^2\Omega A_{rA}^{-1}~.
\label{eq:amloss3}
\end{equation}
At this point, upon assuming a ``dynamo relationship'' $B_{r0}\propto\Omega$ and a fixed moment of
inertia, eq.~(\ref{eq:amloss3}) yields ${\dot \Omega}\propto -\Omega^3$, as required
to match Skumanich's Law, \emph{provided} the Alfv\'en speed $A_{rA}$ at the Alfv\'en radius
remains constant in the late phases of spin-down.

\begin{figure}[t]
\centering
\includegraphics[width=.95\columnwidth]{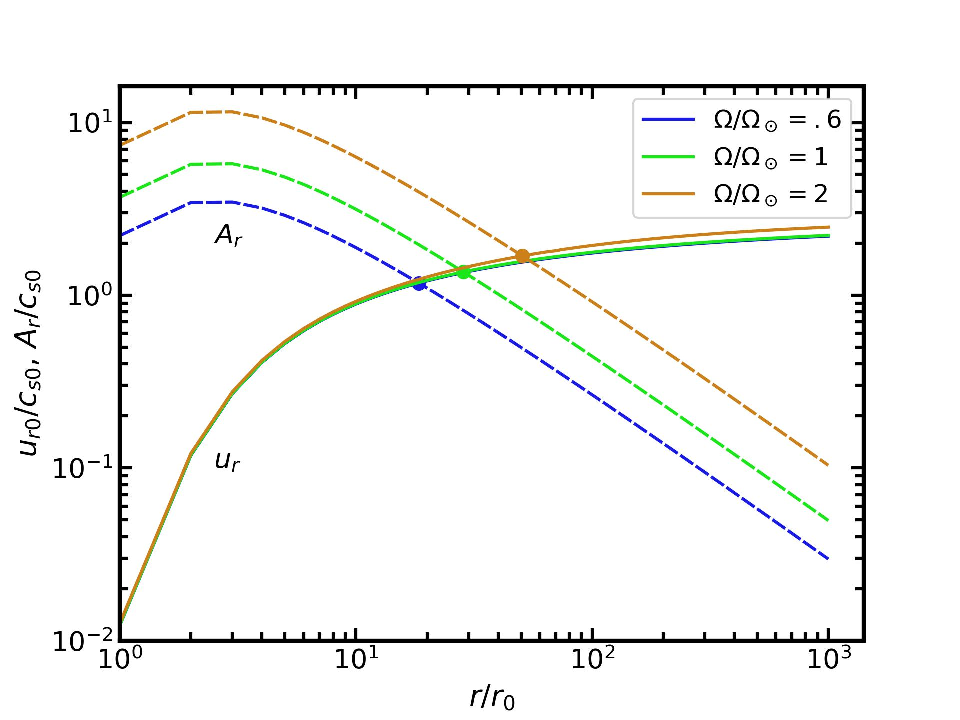}
\caption{
Radial profiles of wind (solid lines) and Alfv\'en (dashed) speed from WD solutions with 
rotation rates $\Omega/\Omega_\odot=0.6$ (blue), 1 (green) and
2 (orange). In all cases $r_0/R_\odot=1.15$, $T_0=1.5\times 10^6\,$K, $\alpha=1.1$,
and $B_{r0}\propto\Omega$.
}
\label{fig:WDsolns}
\end{figure}

Figure \ref{fig:WDsolns} shows radial profiles of the wind and Alfv\'en speeds
for three WD solutions, all with $\alpha=1.1$, $T_0=1.5\times 10^6\,$K and
$B_{r0}\propto\Omega$, differing only in their rotation rates, as color-coded.
%
%\begin{figure}[t]
%\begin{center}
%%\epsfbox{Wind_Alfven_speedv2.eps}
%\end{center}
%\epsfxsize=4.0truein
%\epsfbox{Arra_vs_omega_06_2.eps}
%\end{center}
%\caption{
%Radial profiles
%of wind (solid lines) and Alfv\'en (dashed) speed from WD solutions with 
%rotation rates $\Omega/\Omega_\odot=0.6$ (blue), 1 (green) and
%2 (orange). In all cases $r_0/R_\odot=1.15$, $T_0=1.5\times 10^6\,$K, $\alpha=1.1$,
%and $B_{r0}\propto\Omega$.
%Panel B shows the variation of the Alfv\'en speed at the Alfv\'en radius $r_A$
%as a function fo rotation rate, which at low rotation is well fit by a power law
%$A_{rA}\propto\Omega^{0.287}$ (see text).
%}
%\label{fig:WDsolns}
%\end{figure}
Plasma velocity and Alfv\'en speed components are all normalized to the base sound speed 
$c_{s0}=\sqrt{\alpha p_0 /\rho_0}=165\,$km s$^{-1}$ here.
The wind speed profiles are almost indistinguishable, a direct reflection of the wind being almost entirely thermally-driven. As $\Omega$ varies, so does $B_{r0}$, so that the Alfv\'en speed profiles shift vertically by an amount directly
proportional to $B_{r0}$ ---and thus $\Omega$. As a result the Alfv\'en radius
$r_A$, where solid and dashed curves cross, moves inwards as $\Omega$ (and $B_{r0}$) decreases,
with a concomitant decrease of the Alfv\'en speed $A_{rA}$ at that point. 
%Figure \ref{fig:WDsolns}B shows
From such wind solutions the variation of $A_{rA}$ with $\Omega$ can me measured, and turns out to be well fit by a power law:
\begin{equation}
{A_{rA}(\Omega)\over c_{s0}}=1.348\,\left({\Omega\over\Omega_\odot}\right)^{0.29}~,\qquad 0.6\leq \Omega/\Omega_\odot\leq 2~.
\end{equation}
Again at fixed moment of inertia, Eq.~(\ref{eq:amloss3}) would then ``predict''
$\Omega\propto t^{-0.58}$, a little steeper than Skumanich's Law.

Many options are available to recover more precisely Skumanich's $t^{-1/2}$ Law, for example here assuming a
variation of the coronal base density 
$\rho_0\propto\Omega^{0.29}$.
This type of ad hoc fix is in fact common practice in many semi-empirical angular momentum loss 
models, in which power-law relationships 
of the form ${\dot M}\propto\Omega^a$ and/or $B_{r0}\propto \Omega^b$ are introduced, with the exponents $a,b$ adjusted
to recover ${\dot J}\propto -\Omega^3$ given other model ingredients
\citep[see, e.g.,][and \citealt{Skumanich2019} for a critical synthesis of these approaches]{Kawaler1988,Mattetal2012,Ahuiretal2020}.

\subsection{The MacGregor-Brenner spin-down model\label{ssec:spindown}}

%\textcolor{blue}{[This subsection could be greatly shortened in the "real paper"]}
The torque exerted by a magnetized wind on the photosphere can be presumed to be
efficiently transmitted throughout the convective envelope by turbulent stresses. 
Consequently, the wind-mediated torque first spins down the convective envelope, leaving
underneath a more rapidly rotating radiative core, a process known as core-envelope decoupling.
As the spin-down torque decreases rapidly with the decreasing rotation rate of the envelope,
a flux of angular momentum from the core into the overlying convective envelope
slows down the former and mitigates the spindown of the latter
until, by the solar age, the radiative core spins as a solid body, at the same rate 
as the average rotation rate of the envelope \citep{Tomczyketal1995}.

In a G-type main-sequence star like the Sun, the convective envelope 
($0.713\leq r/R_\odot\leq 1$) accounts for
$\simeq 1$\% of the stellar mass but $\simeq 10$\% of the stellar moment of inertia.
\cite{MacGregorBrenner1991}
have developed a simple two-zone model capturing this process of core-envelope
decoupling and later recoupling, involving a single adjustable parameter, namely
a coupling timescale $\tau_c$ for the exchange of angular momentum between the core and envelope,
each rotating as solid bodies but not necessarily at the same angular velocity.
For fixed moments of inertia for the core ($I_c$) and envelope ($I_e$)
on the ZAMS, the governing equations can be expressed as 
\begin{eqnarray}
\label{eq:MBOc}
	I_c\dert{\Omega_c}{t}&=&-{\Delta J\over\tau_c}~,\\
	I_e\dert{\Omega_e}{t}&=&{\Delta J\over\tau_c}+{\dot J}~,
\label{eq:MBOe}
\end{eqnarray}
where ${\dot J}$ ($<0$) is the angular momentum loss rate in the magnetized wind, as given
by eq.~(\ref{eq:amloss3}), and
\begin{equation}
\Delta J={I_eJ_c-I_cJ_e\over I_c+I_e}={I_cI_e\over I_c+I_e}(\Omega_c-\Omega_e)~.
\end{equation}
The observed rotational evolution
of late-type stars in the age range
$10^8-10^9\,$yr offer a relatively stringent constraint on the value of the timescale
$\tau_c$, which must be  few tens of Myr in order to reproduce the rapid spin down 
of young solar-type stars 
in their first $\simeq 10^8\,$yr on the main sequence \citep{MacGregorBrenner1991,Denissenkovetal2010,GalletBouvier2013}. 
Such a coupling timescale
also allows ``recoupling'' at later times, during which the transfer of angular momentum from the core to the envelope offsets in part loss in the magnetized wind, which provides a natural explanation
for the phase of ``stalled'' spin-down characterizing late-type stars in the 
1--2 Gyr age range \citep{GalletBouvier2013,Curtisetal2020,Gordonetal2021}. 
A good fit to rotational data does require the coupling timescale $\tau_c$ to be a function of stellar mass \citep[see][]{SpadaLanzafame2020}.

The physical mechanism governing angular momentum exchange between the core and envelope,
in principle setting the value of the coupling timescale $\tau_c$,
is not yet identified with confidence. 
Many plausible options are known and documented, 
ranging from hydrodynamical shear instabilities \citep{Zahn1992},
gravity waves \citep{TalonCharbonnel2005,Denissenkovetal2008},
turbulence and magnetic stresses
driven by MHD instabilities \citep{BalbusHawley1994,Spruit2002},
and/or magnetic stresses from a weak internal fossil magnetic field 
\citep{MestelWeiss1987,CharbonneauMacGregor1993,RuedigerKitchatinov1996}. 

%In what follows we adopt the \cite{MacGregorBrenner1991} spin-down model to evolve our model stars to the solar age and beyond,
%using a value of $\tau_c$ within the range allowed by the aforementioned observational constraints.

\begin{figure}[t]
\centering
\includegraphics[width=.95\columnwidth]{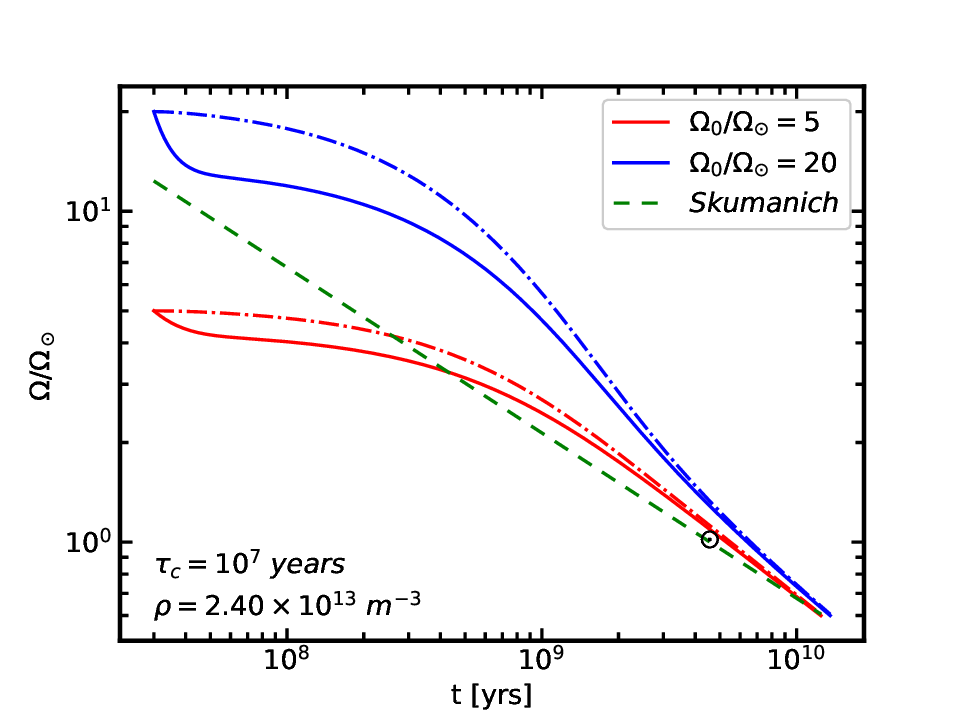}
\caption{ Two representative simulation runs of
the \cite{MacGregorBrenner1991} spindown model, for fixed coupling timescale
$\tau_c=10^7\,$yr and initial conditions on the ZAMS $\Omega/\Omega_\odot=5$ and $25$. Solid (dash-dotted) lines give the angular velocities of the convective envelope (radiative core) in each case.
The dashed line is the Skumanich
$t^{-1/2}$ Law pinned to the present day sun.   }
\label{fig:MacGBsolns}
\end{figure}

Figure \ref{fig:MacGBsolns} shows two representative \cite{MacGregorBrenner1991}
rotational evolution simulations, both using a core-envelope coupling timescale $\tau=10\,$Myr and wind torque 
given by the WD model of the preceding section ($T_0=1.5\times 10^6\,$K, $\alpha=1.1$, $B_{r0}\propto\Omega_e$). The various
curves show the time-evolution of the angular velocity for the radiative core $\Omega_c$
(dash-dotted lines) and convective envelope $\Omega_e$ (solid lines), 
starting here from solid-body rotation
$\Omega_c=\Omega_e$ on the ZAMS, at either 20 (blue) or 5 times (red) the present solar rotation
rate $\Omega_\odot=2.67\times 10^{-6}\,$rad s$^{-1}$. This initial condition is unrealistic, but has no significant influence on the late rotational evolution, on which our subsequent analyses focus.

%\begin{figure}[t]
%\begin{center}
%\epsfxsize=3.5truein
%\epsfbox{M-B_demo.eps}
%\end{center}
%\caption{
%Two representative simulation runs of
%the \cite{MacGregorBrenner1991} spindown model, for fixed coupling timescale
%$\tau_c=10^7\,$yr and initial conditions on the ZAMS $\Omega/\Omega_\odot=5$ and $25$.
%The dashed line is the Skumanich
%$t^{-1/2}$ Law pinned to the present day sun.   }
%\label{fig:MacGBsolns}
%\end{figure}
%
The rapid early spin-down and ``plateau'' in the $\sim 0.05$--$0.5\,$Gyr range are both
clearly seen, as well as the recoupling at later time and general convergence to 
the same rotational evolution trend beyond a few Gyr. This is simply a consequence of angular momentum loss in the WD magnetized wind solution increasing sharply
with rotation rate, i.e., initially faster rotators spin-down faster, so that memory of the initial condition is eventually lost \citep{Kawaler1988}. 
Here, in this asymptotic
regime, $\Omega_e\propto t^{-0.58}$, as expected given the
rotation rate dependence of the WD torque in the thermally-dominated driving regime, viz.~\S\ref{ssec:WDwind}. 

%\smallskip
%The alternate explanatory framework we explore in this section is a collapse of the coronal wind, specifically its associated mass and angular momentum loss rates, following a gradual decrease in coronal temperature.

\section{Magnetized coronal winds near the hydrostatic limit\label{sec:lowT}}

The mechanism(s) responsible for coronal heating have not yet been identified, but magnetic fields are generally believed to be involved \citep{JudgeIonson2024}, so it is certainly plausible that a drop in surface magnetism due to decreasing rotation may end up reducing the coronal temperature to the point
that a transsonic wind no longer materializes.
It is readily shown that in the $1/r^2$ gravity field exterior to a mass
$M_\odot$, a hydrostatic polytropic corona  of finite radial extent can exist as long as the
base coronal temperature satisfies:
\begin{equation}
\label{eq:hydrostatic}
T_0 \leq {\alpha-1\over\alpha} {GM_\odot\mu m_p\over k_B r_0}~, \qquad 1< \alpha\leq 5/3~.
\end{equation}
where $\mu=0.5$ is the mean molecular weight of the coronal plasma, $\alpha$ is the polytropic index, and other symbols have their usual meaning (see Appendix A.1). In what follows we refer to this limiting temperature as the \emph{hydrostatic limit}. For $\alpha=1.1$, this hydrostatic limit is at $T_0=0.91\times 10^6\,$K.

\begin{figure}[ht]
\centering
\includegraphics[width=.95\columnwidth]{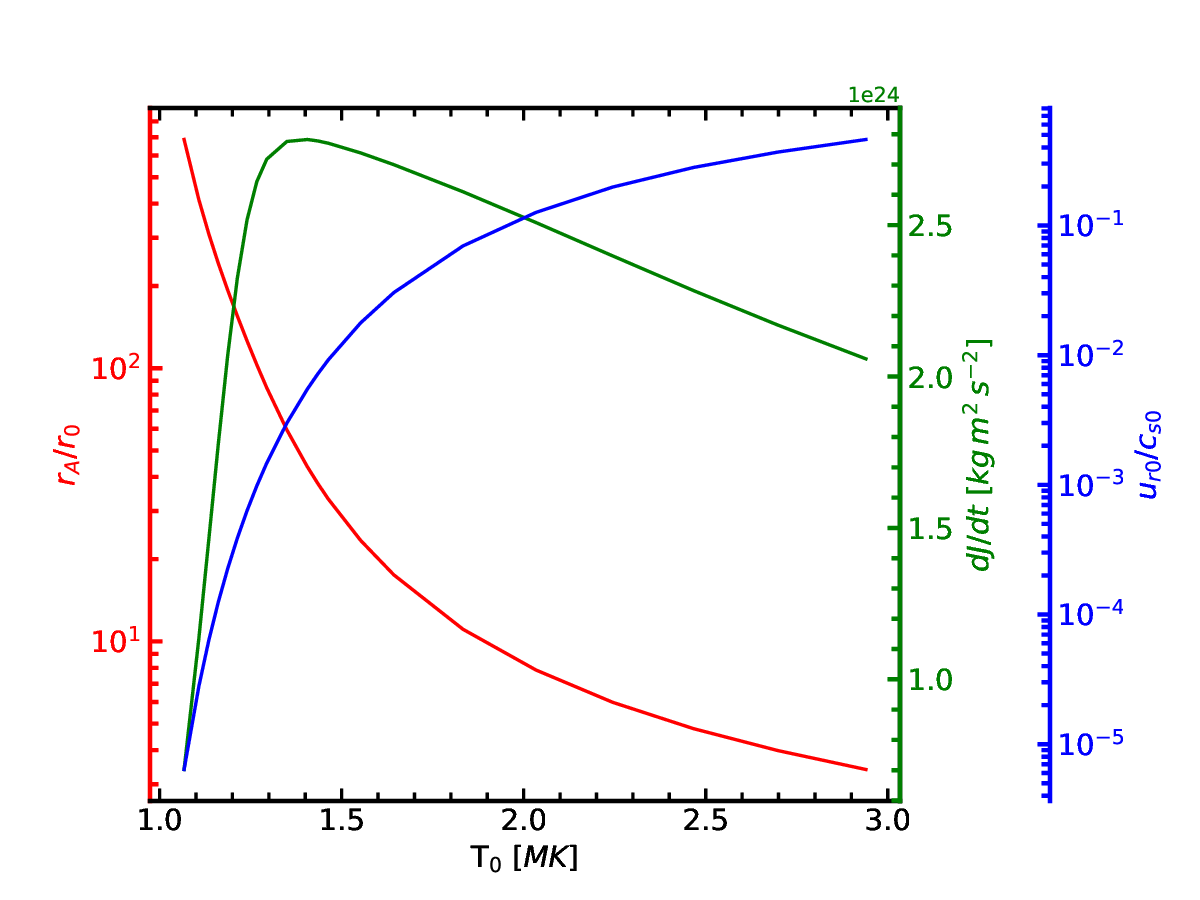}
\caption{
Variations of the Alfv\'en
 radius $r_A$ (red), base flow speed $u_{r0}$ (blue), and
 angular momentum loss rate ${\dot J}$ (green), versus coronal base temperature, as obtained from a sequence of $\alpha=1.1$ WD wind solutions 
 at fixed solar rotation rate and surface magnetic field strength.
}
\label{fig:JdotvsT}
\end{figure}

We begin by investigating
how wind dynamics and associated mass and angular momentum loss rates change as the coronal temperature
decreases towards the limiting value given by Eq.~(\ref{eq:hydrostatic}).
 A sequence of WD wind solutions is computed for base coronal temperatures in the range $1.125\leq T_0/10^6\,{\rm K}\leq 2.93$, using a fixed polytropic index $\alpha=1.1$, with rotation rate and average surface field strength fixed at their solar values, $\Omega_\odot=2.67\times 10^{-6}\,$rad s$^{-1}$ and $B_{r0}=2.79\times 10^{-4}\,$T ($2.79$ G) in all cases. Figure \ref{fig:JdotvsT} shows the variations of the Alfv\'en
 radius $r_A$, base flow speed $u_{r0}$, and
 angular momentum loss rate ${\dot J}$, as a function of coronal base temperature $T_0$.
Moving from right to left,
as $T_0$ drops from 2 to $1.3\times 10^6\,$K, the
base wind speed $u_{r0}$ (blue curve), 
and thus the mass loss rate, drops rapidly; however the Alfv\'en radius $r_A$ (red curve) increases markedly, a direct consequence
of the decreasing wind speed while the Alfv\'en speed radial profile remains fixed since here both the base field strength and density are held constant
(viz.~Fig.~\ref{fig:WDsolns}). This more than compensates the drop in mass loss rate, with the consequence that the angular momentum loss rate
${\dot J}$ (green curve, as computed via Eq.~(\ref{eq:amloss3}) \emph{increases}. This behavior is not specific to the WD wind model, and has been
noted in geometrically more complex simulations \citep[see, e.g., the 2D MHD polytropic wind simulations of][]
{PantolmosMatt2017}. However, once $T_0$
drops below $\simeq 1.45\times 10^6\,$K the precipitous drop in base flow
speed now surpasses the increase in $r_A^2$, so that ${\dot J}$ now \emph{decreases} as $T_0$ is
further reduced. The angular momentum loss rate is
thus a non-monotonic function of base coronal temperature,
peaking here at $T_0\simeq 1.35\times 10^6\,$K. 

\begin{figure}
\centering
\includegraphics[width=.95\columnwidth]{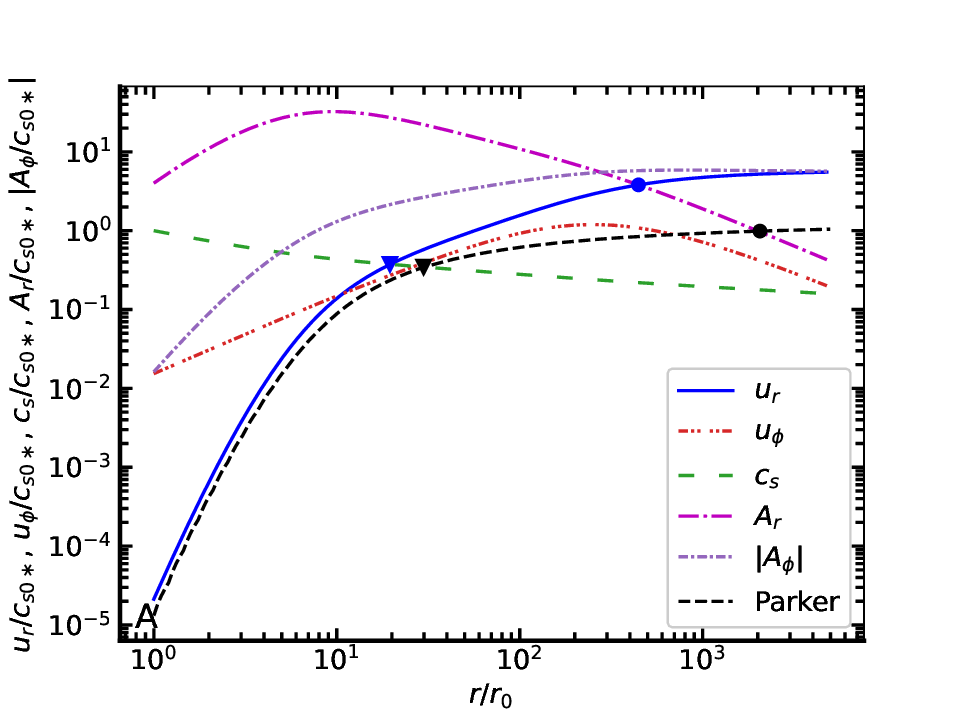}
\includegraphics[width=.95\columnwidth]{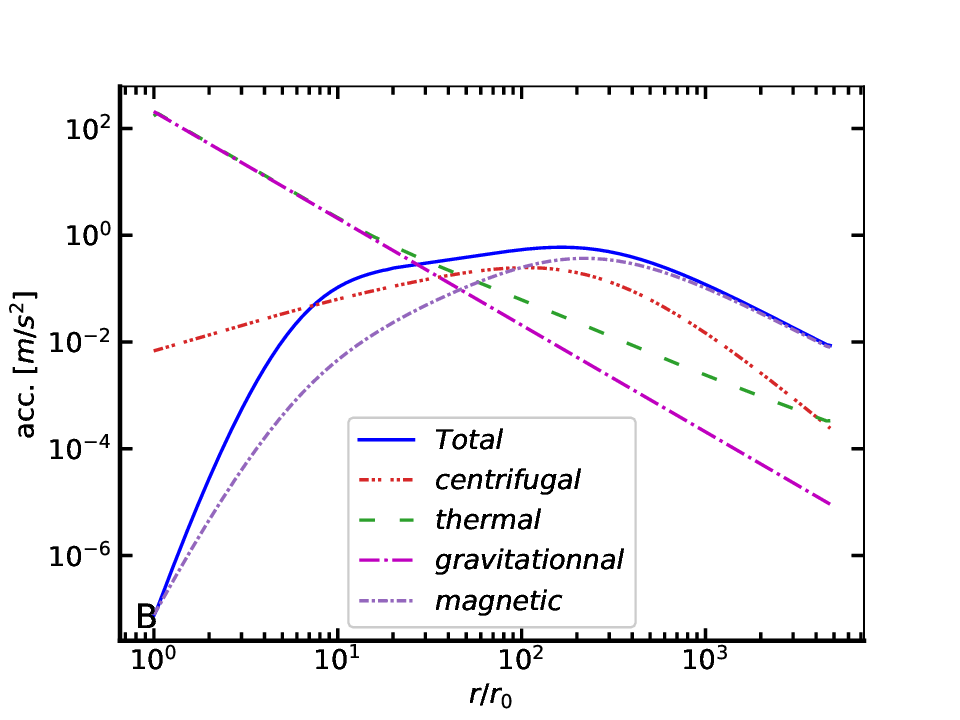}
\caption{
Top panel: Radial profiles of wind variables for a $\alpha=1.125$
WD solution with solar rotation and magnetic field strength, and a ``low'' coronal base temperature $T_0=1.23\times 10^6\,$K. 
The black dashed line shows the $u_r(r)$ profile for a non-rotating, unmagnetized coronal wind of the same base temperature and polytropic index. The two solid dot indicates the corresponding Alfv\'en points.
The bottom panel shows the corresponding acceleration terms in the momentum equation: gravity, pressure, centrifugal, and magnetic force, as labeled. The solid blue line gives the resulting Lagrangian acceleration. Note how centrifugal and magnetic forces dominate beyond the sonic point (blue inverted triangle).
}
\label{fig:WDlowT}
\end{figure}

The non-monotonic variation of ${\dot J}$ with coronal base temperature reflects a change in dynamical regime taking place in the winds emanating from coronae with decreasing base temperatures.
For a slowly rotating, weakly-magnetized solar-type star, once the coronal temperature approaches
the hydrostatic limit (\ref{eq:hydrostatic}) from above, magnetocentrifugal driving, negligible for a 
$1.5\times 10^6\,$K corona with $\alpha=1.1$, becomes significant, as it would be
in young, rapidly rotating and strongly magnetized stars of solar coronal temperatures. Figure \ref{fig:WDlowT} shows a
specific example, namely a WD wind solution for
a star with solar rotation rate and surface field
strength, but a coronal temperature of $1.23\times 10^6\,$K and polytropic index
$\alpha=1.125$. The top panel shows radial profile of wind and Alfv\'en speed components, while the bottom panel depicts the corresponding radial force budget as a function of $r$.
%
%\begin{figure}
%\begin{center}
%\epsfxsize=3.5truein
%\epsfbox{alpha_1.125_T_0.82.eps}
%\epsfxsize=3.5truein
%\epsfbox{Force.eps}
%\end{center}
%\caption{
%Top panel: radial profiles of plasma variables for a $\alpha=1.125$
%WD wind solution
%with solar rotation and magnetic field strength, and a ``low'' coronal base temperature $T_0=1.23\times 10^6\,$K. The solid dot indicates the sonic point.
%The black dashed line shows the $u_r(r)$ profile for a non-rotating, unmagnetized coronal wind of the same base temperature and polytropic index.
%The bottom panel shows the corresponding acceleration terms in the momentum equation: gravity, pressure, centrifugal, and magnetic force, as labeled, and the solid blue line gives the resulting Lagrangian acceleration. Note how centrifugal and magnetic forces dominate beyond the sonic point.
%}
%\label{fig:WDlowT}
%\end{figure}

Because the thermal pressure force is much reduced due to the low coronal temperature, centrifugal and magnetic forces, even if small in absolute terms, dominate the wind dynamics beyond the sonic point 
(the radius at which $u_r=c_s$, indicated by inverted triangles on
Fig.~\ref{fig:WDlowT}), leading to an asymptotic flow speed $858\,$km s$^{-1}$, significantly larger than the $162\,$km s$^{-1}$  characterizing a purely hydrodynamical coronal wind \`a la \cite{Parker1958} (dashed black line on Fig.~\ref{fig:WDlowT}A) at the same coronal temperature. However, this increase in asymptotic wind speed does not translate into an enhanced mass loss rate, the base flow speed being essentially identical for the WD and Parker wind solutions; at large distances, the WD wind simply has a faster speed but lower density. This is a well known property of coronal winds, independent of the mechanism depositing energy or angular momentum in the wind, over and above thermal driving
\citep{LeerHolzer1980,Ofman2010}. From the point of view of angular momentum loss, the important take-home message is
that for such low base temperature winds,
while magnetocentrifugal driving does not greatly affect the base flow speed, it does displace the Alfv\'en radius significantly inwards, thus further \emph{reducing} the angular momentum loss rate as compared to a
naive estimate using the wind profile $u_r(r),\rho(r)$ of the Parker unmagnetized wind model to estimate the location of the Alfv\'en radius, cf.~the two dots on Fig.~\ref{fig:WDlowT}A.
%wind dominate by the pressure gradient represented by the Parker model on Figure \ref{fig:WDlowT}A.}

\begin{figure}
\centering
\includegraphics[width=.95\columnwidth]{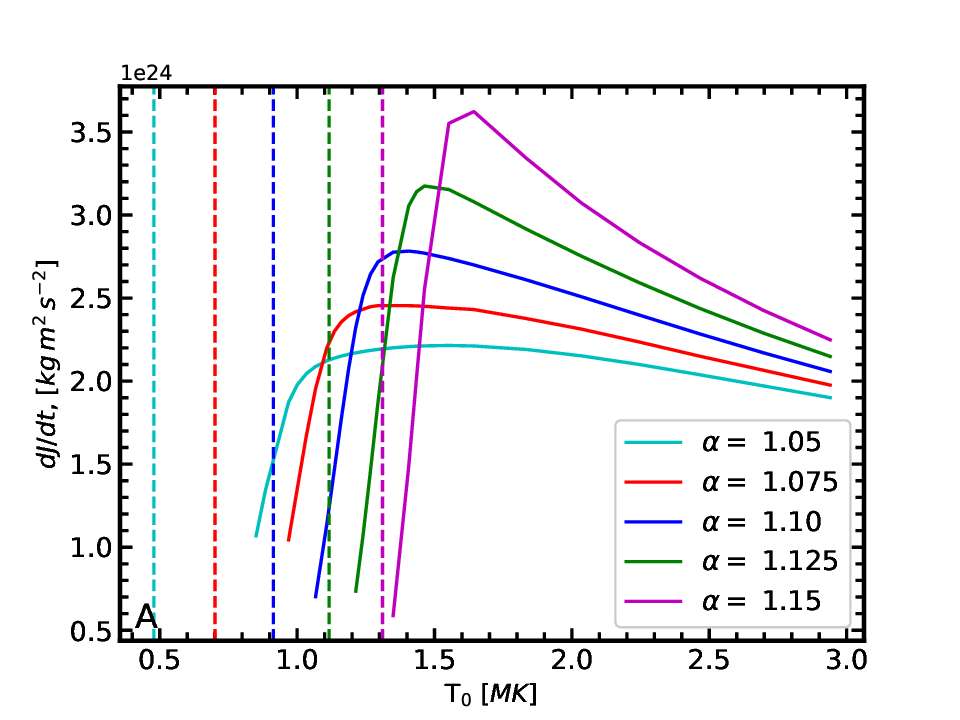}
\includegraphics[width=.95\columnwidth]{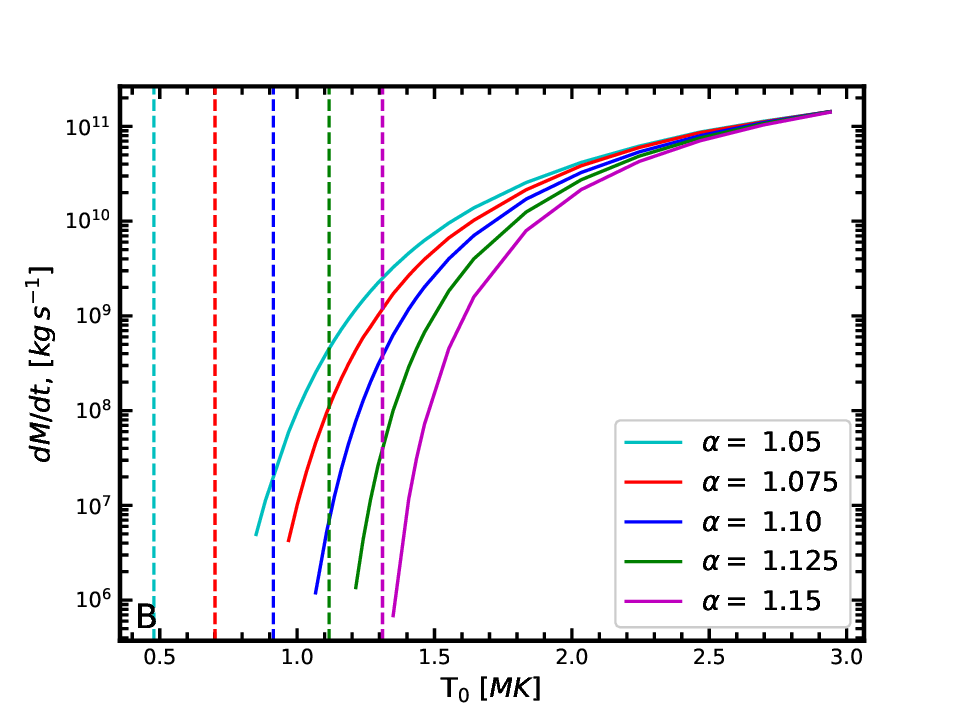}
\caption{Angular momentum loss rate (top panel) and
mass loss rate (bottom panel) as a function of coronal base temperature
in WD wind solutions with various values of the polytropic index $\alpha$, as color-coded. The vertical dashed lines indicate the corresponding hydrostatic corona limiting temperature. In all cases the rotation rate and magnetic field strength are held at their solar values, and the coronal base density is fixed at $\rho_0=10^{14} m_p\,$m$^{-3}$ in calculating the mass loss rate. 
}
\label{fig:AM+Mloss}
\end{figure}

The consequence of this complex dynamical behavior is that,
as shown
on Figure \ref{fig:AM+Mloss},
the angular momentum loss rate ends up varying in a non-monotonic manner with both the base temperature and polytropic index, the latter being another important parameter ultimately related to coronal heating (more on this in \S\ref{sec:Wenergetics} below).
Note how at the higher base temperatures,
the solutions with polytropic index closer to adiabatic have higher angular momentum loss rates, but wind solutions closer to isothermal ($\alpha\to 1.0$) retain an approximately constant angular momentum loss rate to
much lower base temperatures before wind collapse occurs. 
Mass loss (Fig.~\ref{fig:AM+Mloss}B), on the other hand, decreases monotonically with decreasing coronal base temperature $T_0$ and increasing polytropic index $\alpha$, the collapse
of mass loss taking place at a base temperature decreasing with decreasing $\alpha$. We note also that the decrease of mass loss with base temperature, although monotonic, cannot be well fit by a power-law in this temperature range, no matter the value of $\alpha$.

The abruptness seen on Fig.~\ref{fig:AM+Mloss} in the collapse of the angular momentum loss rate as coronal temperature drops over the peak in ${\dot J}$ is a direct consequence of magnetocentrifugal driving: as shown on
Fig.~\ref{fig:WDlowT}, at low coronal base temperatures it increases the flow speed beyond the slow magnetosonic point,
which greatly reduces the outward displacement of the Alfv\'en radius normally associated with decreasing overall wind speeds (cf.~Fig.~\ref{fig:JdotvsT}), so that it can no longer compensate the decrease in mass loss rate associated with a lower base flow speed, as it does at higher coronal temperatures. At the dynamical level, magnetocentrifugal driving becomes important at low coronal temperature because, as $u_r$ drops, more winding of the magnetic field can take places, which increases the strength of the azimuthal magnetic component in the wind per unit radial displacement of a plasma volume element. Consequently, both the radial and azimuthal components of the Lorentz force also becomes larger.

 Fig.~\ref{fig:AM+Mloss} already indicates that there are multiple possible pathways to the shutdown of angular momentum loss. In addition to the (obvious) decrease in base coronal temperature,
close examination of Figure \ref{fig:AM+Mloss}A reveals that in the
coronal temperature range $0.8\leq T_0\leq 1.5$MK, at fixed $T_0$ a gradual increase in the polytropic index ---amounting to reduced power input into the corona, as detailed in \S\ref{sec:Wenergetics} below--- would generate a moderate rise followed by an extremely swift drop
of the angular momentum loss rate.

\section{Breaking Skumanich's Law\label{sec:breakSku}}

\begin{figure}
\centering
\includegraphics[width=.82\columnwidth]{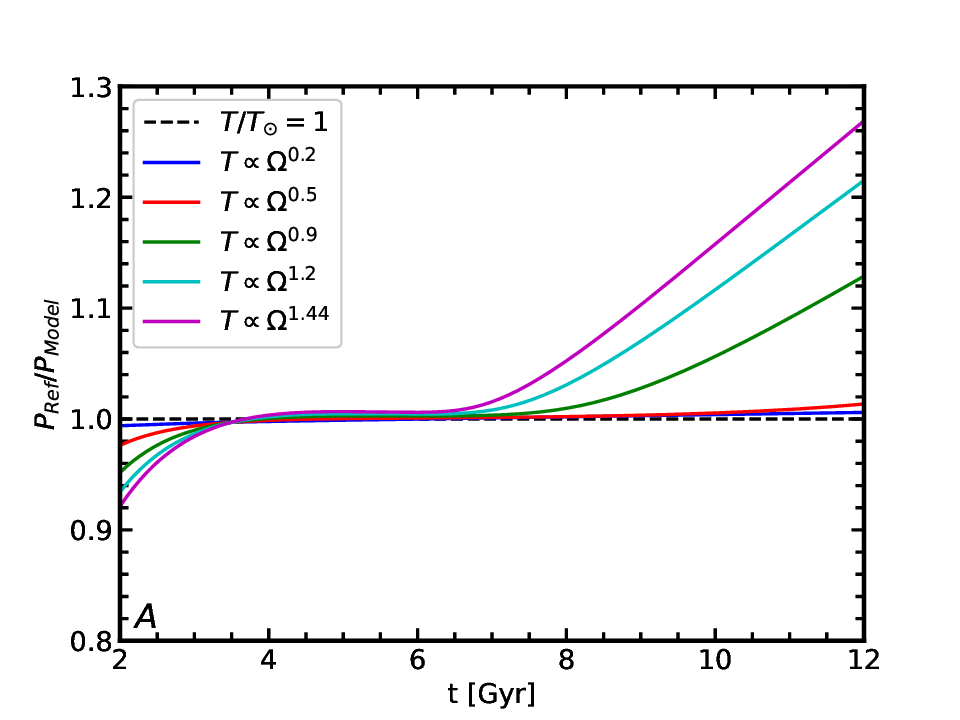}
\includegraphics[width=.82\columnwidth]{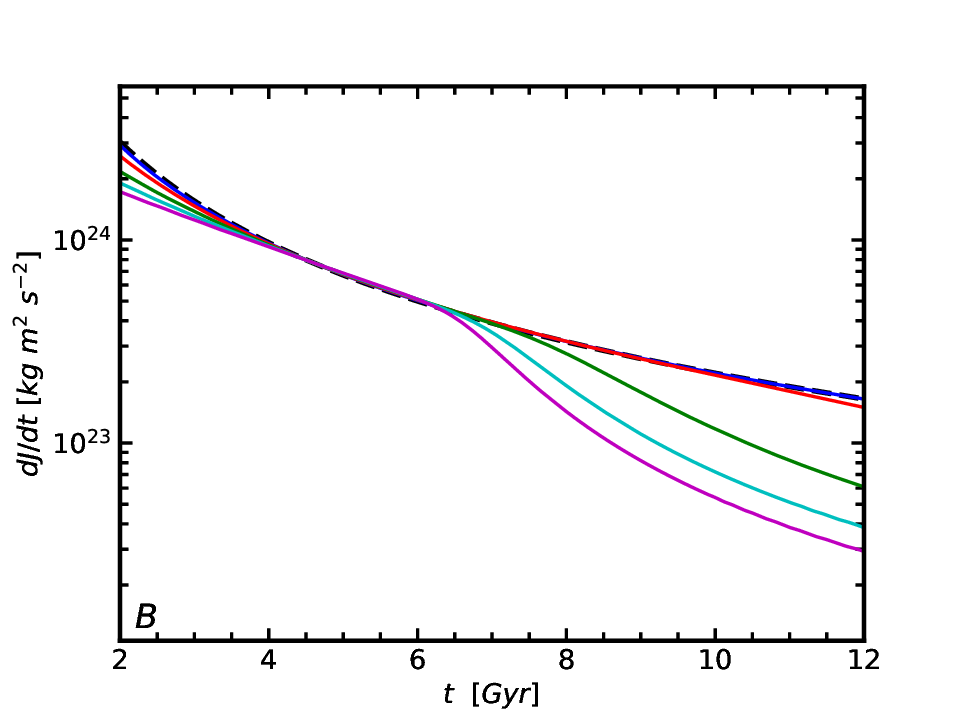}
\includegraphics[width=.82\columnwidth]{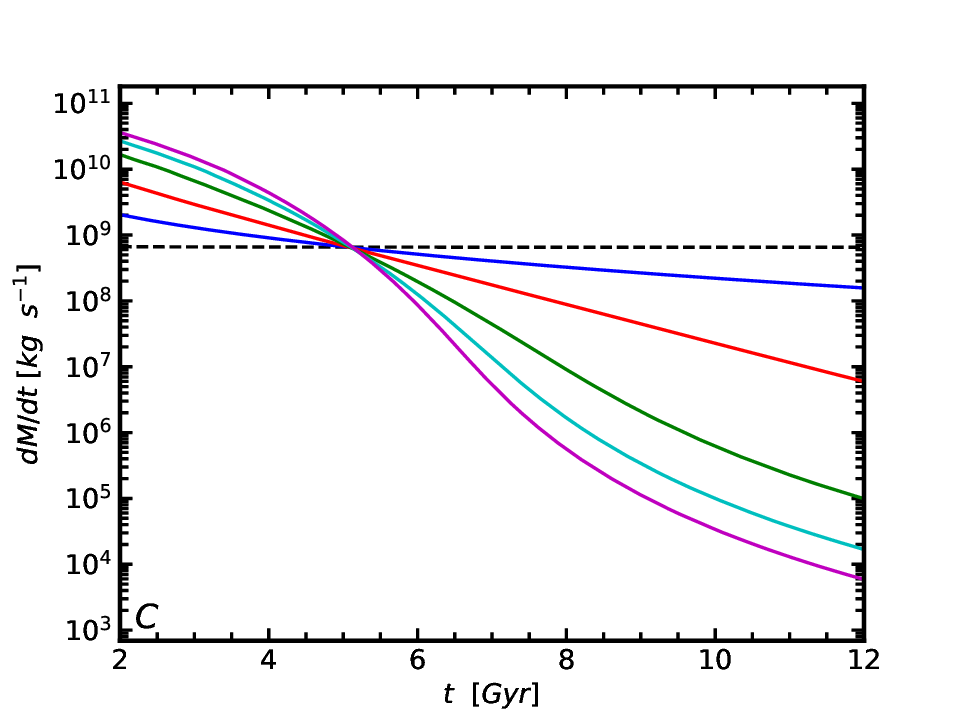}
\includegraphics[width=.82\columnwidth]{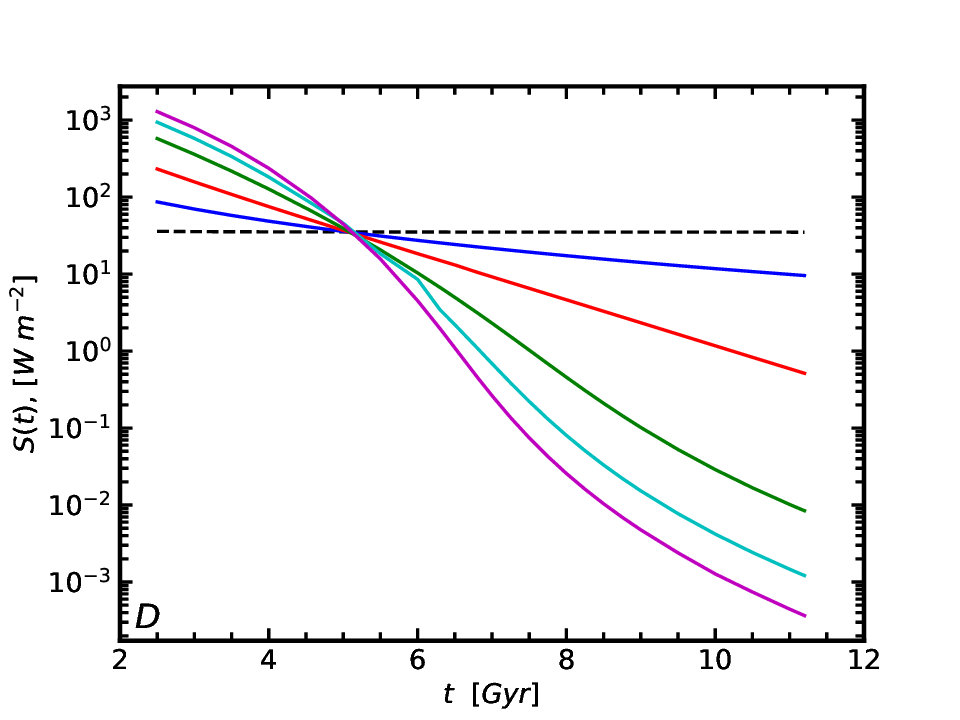}
\caption{
 Break of the Skumanich Law from our rotational evolution model using different values of $\sigma$  in equation \ref{eq:TempvsOmega_sigma}. The figures show the evolution of (A) the ratio $P_{\rm Ref}/P_{\sigma}$ of rotation periods, (B) ${\dot J}$, (C) ${\dot M}$ and (D) $S$ as a function of time. Here $P_{\rm Ref}$ is the rotation period for a model for which temperature is fixed and $P_{\sigma}$ is the period for models where the temperature evolves as described by equation \ref{eq:TempvsOmega_sigma} and $S$ is the integrated volumetric heating rates as described by equation (\ref{eq:heatsource}). The same colour coding applies to all panels.}
\label{fig:BreakSku}
\end{figure}

Guided by the modelling results presented in the preceding section,
we now investigate the conditions under which Skumanich's Law can be broken (or not) by collapse of the coronal wind triggered by a gradual reduction of the coronal base temperature
as the star spins down.
To do so, and following \cite{OFionnagainVidotto2018}, we 
introduce an ad hoc power-law relationship between base coronal temperature and surface magnetic field $B_{r0}$ which, because of our assumed dynamo relationship $B_{r0}\propto \Omega$ then translates into a power-law relationship with rotation rate:
\begin{equation}\label{eq:TempvsOmega_sigma}
{T_0(\Omega)\over T_{\odot}}
=\left({\Omega\over\Omega_\odot}\right)^\sigma~,\qquad\sigma\geq 0~.
\end{equation}
with $T_\odot=1.5\times 10^6\,$K.
Figure \ref{fig:BreakSku}A shows the rotational evolution resulting from this Ansatz, for varying values of the exponent $\sigma$, as labeled. What is plotted is the ratio of the rotation period to that of a reference run with constant $T_0=1.5$MK. All solutions use the dynamo relationship
$B_{r0}\propto\Omega$ and $\alpha=1.1$, and are computed using the \cite{MacGregorBrenner1991} model.  The initial conditions are set so that the angular momentum loss is fixed at the solar value when the model reaches the age of the Sun.
%Runs spin-down avec $B_{r0}\propto\Omega$ et $\alpha=1.1$ comme d'habitude, debutant a un age de $\simeq 3\,$Gyr et/ou
%$\Omega/\Omega_\odot=2$, force a passer par $\Omega/\Omega_\odot=1$ à 4.5Gyr (par decalage de l'echelle de temps). Condition initiale $\Omega_c$, $\Omega_e$ provenant d'une run MB a $\tau_c=10\,$Myr ? Une solution de reference "Skumanich" avec
%
In light of Fig.~\ref{fig:AM+Mloss}A, one might expect that a drop of $\simeq 25\%$ in temperature is required to shut off angular momentum loss, while the Skumanich relation would predict a drop of $\simeq 20\,$\% in rotation rate; this would then require $\sigma\simeq 1$ to achieved a significant reduction of angular momentum loss between 5 and 8 Gyr. This expectation is consistent with the results plotted on
Fig.~\ref{fig:BreakSku}A, but even at $\sigma=1.44$ the deviation from Skumanich's Law at 8 Gyr is modest, with
only a $\simeq 10\,$\% increase in rotation period.

Figure \ref{fig:BreakSku} also reveals that despite a drop by many orders of magnitude in the mass loss rate beyond the solar age for
evolutionary sequences with $\sigma \geq 0.5$ (panel C), the corresponding decrease in angular momentum loss rate (panel B) remains quite modest.
This is a direct consequence of the peculiar dynamics of magnetized coronal winds at temperatures approaching the hydrostatic corona limit, as detailed in \S\ref{sec:lowT}. In this regime, mass loss is emphatically not a good predictor of angular momentum loss.

\section{Wind energetics\label{sec:Wenergetics}}

The energetics of coronal wind acceleration is inseparable from the coronal heating problem \citep{HansteenLeer1995}. In regions of the corona overlying the quiet sun, it is estimated that up to 17\% of the $\simeq 800\,$W m$^{-2}$ energy flux required to maintain the corona in a quasi-steady state is used in accelerating the solar wind; this fraction goes up to $\simeq 90\,$\% in coronal holes from which the fast component of the solar wind originates
\citep[see discussion in][\S 1.5 and Appendix B]{JudgeIonson2024}

While a quasi-consensus exists to the effect that magnetic fields are involved at some level,
the physical mechanism(s) responsible for coronal heating is yet to be identified with confidence. The collection of review papers introduced by
\cite{DeMoortelBrowning2015} offer a nice sample of current ideas \citep[see also][]{Judge2023,Judgeetal2024}. If the observed break of gyrochronology in solar-type stars older than the sun is indeed associated with thermal collapse of coronal winds, then useful constraints regarding coronal heating may perhaps be derived from spindown considerations.

In wind models (including the WD model) or MHD numerical simulations relying on the polytropic approximation $p\propto\rho^\alpha$, the coronal base temperature and polytropic index are typically input parameters which can be varied independently; in fact, and as detailed in what follows, the polytropic index embodies a specific
form of volumetric heating profile; 
$\alpha=1$ corresponds to an isothermal corona, and thus an (unrealistic) infinite energy source, while $\alpha=5/3$ corresponds to adiabatic expansion, i.e., vanishing volumetric energy input
\citep[see, e.g., \S 4.2 in][]{LamersCassinelli1999}.
%, in which case no steady wind solution is possible. \textcolor{red}{[CHECK VERSUS EQ.~(\ref{eq:hydrostatic})]}
The coronal base temperature, on the other hand, is a \emph{result}
of energy balance between coronal heating, and energy lost to radiation, thermal conduction, and acceleration of the solar wind.

\begin{figure}
\centering
\includegraphics[width=.95\columnwidth]{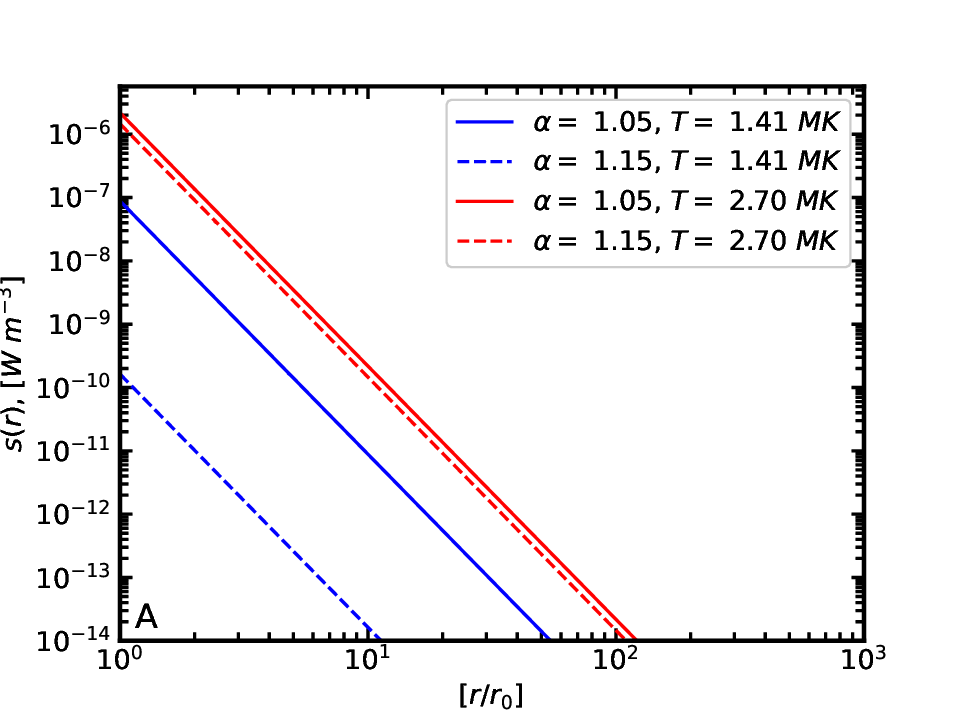}
\includegraphics[width=.95\columnwidth]{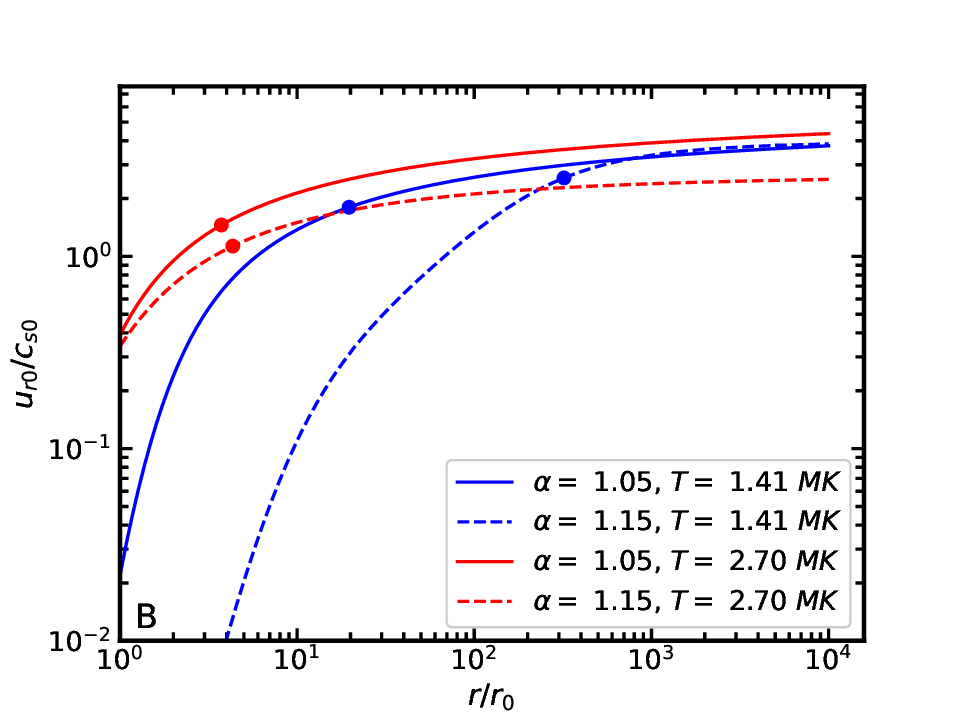}
\caption{Panel A shows radial profiles
of volumetric power input in a sample of polytropic WD wind solutions, reconstructed a posteriori via Eq.~(\ref{eq:heatsource}). Panel B shows the corresponding flow profiles. All solutions are computed for solar rotation and surface magnetic field strength. The solid dots indicate the position of the Alfv\'en radius $r_A$ in each solution.
}
\label{fig:WDheating}
\end{figure}

With a polytropic wind solution already computed, it is possible to reconstruct a posteriori the profile of volumetric heating that would result in the same wind solution. One simply needs to write the energy equation, including now explicitly a volumetric source term $s(r)$ on the RHS:
\begin{eqnarray}
\label{eq:heatsource}
\nabla\cdot\left[\rho{\bf u}
\left({u^2\over 2}+{3p\over 2\rho}
+{B^2\over 2\rho\mu_0}\right)\right]
+\nabla\cdot(p{\bf u}) \nonumber \\
+{\bf u}\cdot\nabla\Phi
-{1\over\mu_0}{\bf u}\cdot({\bf B}\cdot\nabla){\bf B}=s(r)
%{1\over r^2} \ddt{r} \left[
%r^2\rho u_r\left({1\over 2}{\bf u}^2+%{3\over 2}{p\over\rho}\right)
%\right]+\dert{(u_rp)}{r}-\rho u_r {GM_\odot\over r^2}=s(r)
\end{eqnarray}
Substituting the WD polytropic wind solution $u_r(r)$, $p(r)$, $\rho(r)$, etc. in the LHS then yields $s(r)$ by direct computation. The results of such an exercise are shown on the top panel of Figure
\ref{fig:WDheating}, for a selection
of WD wind solutions computed using distinct combinations of $[\alpha,T_0]$ and otherwise solar parameters. The corresponding wind speed profiles $u_r(r)$ are plotted on the bottom panel.

In all cases the volumetric source term
$s(r)$ can be well approximated by a steep power law in heliocentric radius, i.e.,
$s(r)\propto r^{-\gamma}$, with $\gamma\simeq 4$.
It is physically satisfying that the heating profiles extracted in this manner peak at the base of the corona, and decrease rapidly outwards. Yet the relationship to the
two heating parameters, the base flow speed $T_0$ and polytropic index $\alpha$ is intricate; at $T_0=2.7\,$MK the volumetric heating profiles (in red) are nearly independent of $\alpha$, but at the lower temperature of 1.4$\,$MK differ by over two orders of magnitudes at all radial positions.

Comparing the flow speed profiles for two $T_0=1.4\,$MK WD solutions on Fig.~\ref{fig:WDheating}B (blue curves), note how the asymptotic flow speeds are almost identical, while the base flow speed ---and thus mass loss rate--- is orders of magnitude smaller in the $\alpha=1.15$ solutions than for $\alpha=1.05$. Conversely, and despite nearly identical volumetric heating profiles, the two $T_0=2.7\,$MK wind solutions (in red) have distinctly different acceleration profiles and asymptotic flow speeds, but nearly identical base flow speeds. 
Perhaps counter-intuitively given the wind profiles on Fig.~\ref{fig:WDheating}B, all four of these wind solutions have comparable angular momentum loss rates, ranging from
${\dot J}=-1.48 \times 10^{24}\,$kg m$^2\,$s$^{-2}$  for the $(\alpha,T_0)=(1.15,1.41\,{\rm MK})$ solution (dashed blue), up to
${\dot J}=-2.42 \times 10^{24}\,$kg m$^2\,$s$^{-2}$ $(\alpha,T_0)=(1.15,2.69\,MK)$ (solid blue).

\begin{figure}[t]
\centering
\includegraphics[width=.95\columnwidth]{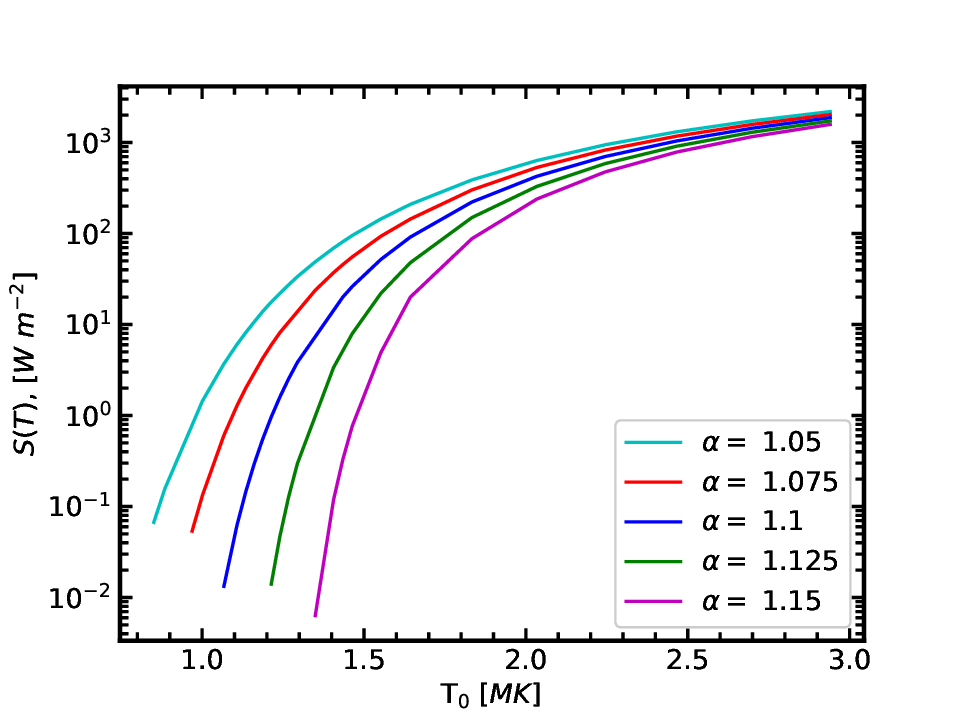}
\caption{Integrated volumetric heating rates
for WD solutions of various $\alpha$ and
$T_0$, expressed as energy flux at the base of the wind, $r_0/R_\odot=1.15$. As on Fig.~\ref{fig:AM+Mloss}, all wind solutions are computed at the solar rotation rate and surface magnetic field strength.
}
\label{fig:WDheatT-alpha}
\end{figure}

Integrating $s(r)$ from the coronal base out to infinity yields the power required to
maintain the coronal temperature and accelerate the wind.
This power requirement, expressed as an energy flux at the base of the wind, is plotted on Figure \ref{fig:WDheatT-alpha}, for sequences of WD solutions at fixed polytropic index $\alpha$ and varying coronal base temperature. 
The total heating rate $S$ drops with decreasing $T_0$ faster than a
power-law, the more swiftly the closer the polytropic index $\alpha$ is to its adiabatic limit $\alpha=5/3$. Below $\simeq 1.5\,MK$, at fixed coronal base temperatures power input decreases rapidly with increasing $\alpha$, but this sensitivity rapidly disappears as $T_0$ increases beyond $\simeq 2\,$MK.
It must be kept in mind that these results pertain to the energy required to propel the wind, and do not include radiative of conductive energy loses. For $\alpha=1.1$, a $T_0=1.5\times 10^6\,$K corona requires $\sim 10^{2}\,$W m${}^{-2}$, in line with observational inferences \citep{JudgeIonson2024}.

\begin{figure}
\centering
\includegraphics[width=.95\columnwidth]{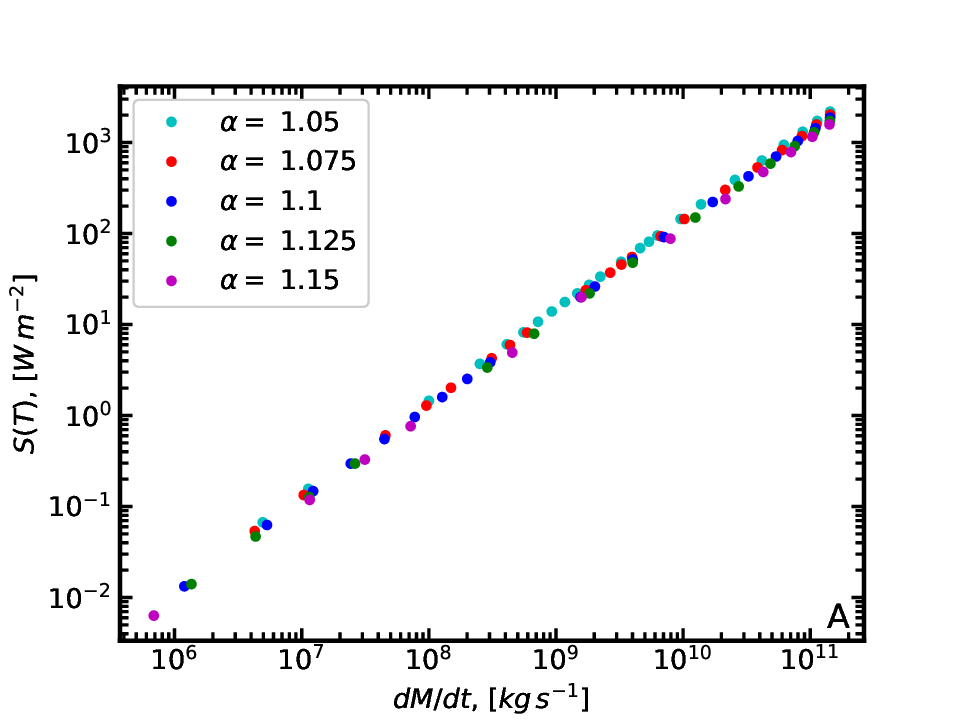}
\includegraphics[width=.95\columnwidth]{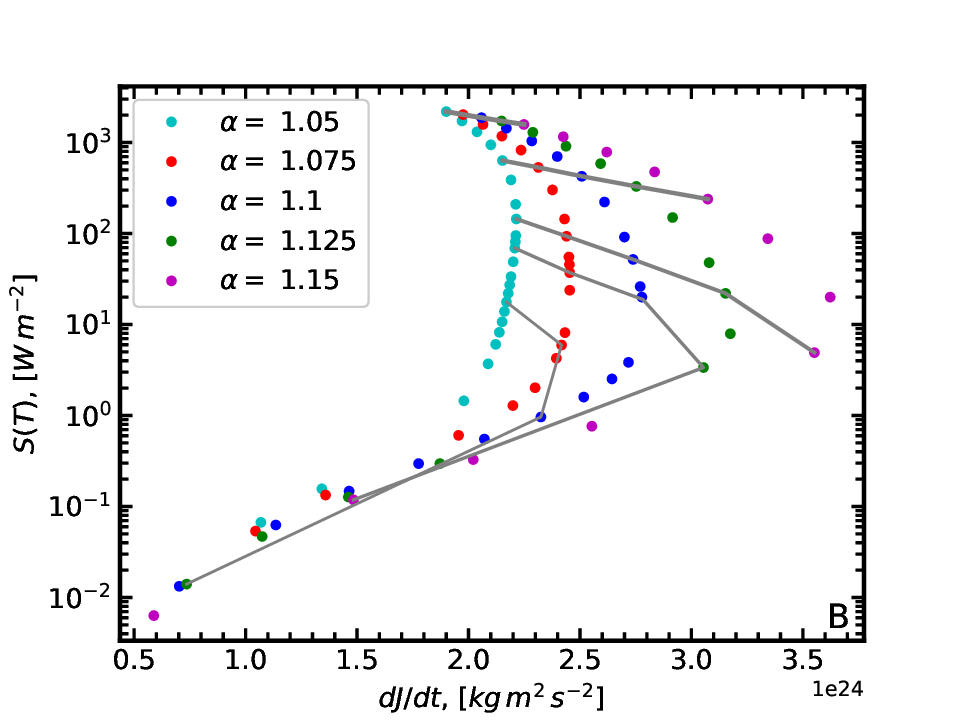}
\caption{Correlation between integrated volumetric heating rates and mass loss rate (top panel) and angular momentum loss rate (bottom panel)
for WD solutions of various $\alpha$ and
$T_0$.  All wind solutions are computed at the solar rotation rate and surface magnetic field strength, and mass loss rates is computed for a base density
$\rho_0=10^{14}m_p\,$m$^{-3}$. The base temperature $T_0$ ranges from $0.85$ to $2.94$ MK and increases upwards along each sequence of colored dots. The lines on panel B are isotherms for values
$T_0=[1.21,1.4,1.55,2.03,2.94]\,$MK, from bottom to top.}
\label{fig:corrWDheat+dmdt}
\end{figure}

Figure \ref{fig:corrWDheat+dmdt} shows the correlation between power input and mass (top panel) and angular momentum (bottom panel) loss rates, for a sample of WD wind solutions taken from Figs.~\ref{fig:AM+Mloss} and \ref{fig:WDheatT-alpha}. Data points are color-coded in terms of the polytropic index $\alpha$, as labeled, and base temperatures increase from bottom to top along each colored-dot sequence.
Although the mass loss rate clearly cannot be described by a power-law in base temperature (viz.~Fig.~\ref{fig:AM+Mloss}B), a single and very tight linear relationship (power-law of index unity) emerges when correlated against
total power input into the wind, characterizing WD wind solutions computed for a wide range of
base temperatures and polytropic indices. The power input is clearly a great predictor of the mass loss rate (and vice-versa!). Note again the marked degeneracies characterizing these WD wind solutions, with widely different
$(\alpha,T_0)$ pairs associated with identical power input, and thus mass loss rate. 
%\textcolor{red}{There is clearly a link to be made with the effect of the centrifugal force on wind acceleration. The effect of the centrifugal and magnetic force clearly explains why the correlation isn't found in $dJ/dt$}

On the other hand, as evidenced on Fig.~\ref{fig:corrWDheat+dmdt}B the angular momentum loss rate shows a more complex relationship to power input, definitely not expressible as a power-law even at fixed $\alpha$. Note that the horizontal axis spans only a factor of $\simeq 4$ in ${\dot J}$, in contrast to the five
orders of magnitude in ${\dot M}$ on panel A; so that in comparison, ${\dot J}$ is almost independent of the power input \emph{despite} the strong dependency of ${\dot M}$.

With these results in hand, we return to the break of Skumanich's Law in the spin-down simulations of \S\ref{sec:breakSku}. Panel D of Figure
\ref{fig:BreakSku} shows the power input
to the wind calculated a posteriori along each spin-down run of panel A. 
As one might have anticipated on the basis of Fig.~\ref{fig:corrWDheat+dmdt}A, the drop in power input along the rotational evolution sequence closely shadows that of the mass loss rate, dropping by some 3 orders of magnitude between 5 and 8 Gyr.
Such a large drop, for which no observational evidence currently exists to the best of our knowledge, represents a strong constraint on any explanation for the break in gyrochronology invoking collapse of mass loss in aged solar-type stars.

\section{Discussion and conclusion\label{sec:conclu}}

%In this paper we have examined if and how a gradual reduction of coronal heating could lead to a sudden and large drop in angular momentum loss rate
%in magnetized winds beyond the solar age, as suggested by the recently observed break in the gyrochronology sequence, otherwise well-captured by the Skumanich Law, $\Omega(t)\propto t^{-1/2}$.

Our investigations into the dynamics of
polytropic magnetized coronal winds near the hydrostatic limit have revealed a  number of somewhat counter-intuitive results, of importance in the context of the spin-down of solar-type stars and observed break of gyrochronology; and raise some warning flags regarding the indiscriminate use of ad hoc power law to relate input wind parameters to mass and/or angular momentum loss rates:

\begin{enumerate}
\item At low ($\lta 1.2\,$MK) coronal base temperatures, magnetocentrifugal driving is important even at the relatively low solar rotation rate ($\Omega$) and coronal magnetic field strength ($B_{r0}$).
It does not greatly affect mass loss,
but can lead to 
 a significant decrease of ${\dot J}$ as compared to pure thermal driving.
 \item In the magnetohydrodynamical Weber-Davis wind models, at fixed rotation rate and magnetic field strength neither the angular momentum loss rate  ${\dot J}$ nor mass loss rate ${\dot M}$ can be described by power-laws in coronal temperature $T_0$; the former is even non-monotonic in $T_0$.
\item Again at fixed $\Omega$ and $B_{r0}$, ${\dot M}$ is entirely set by the power input into the wind, itself
showing strong degeneracies in $(\alpha,T_0)$.
\item As coronal base temperature decreases towards the hydrostatic corona limit, the collapse of ${\dot J}$ is very sudden and takes place at significantly larger $T_0$ than the hydrostatic limit.
\end{enumerate}

We have then attempted to break the Skumanich
$t^{-1/2}$ Law for solar-type stars older than the sun, by introducing
an ad hoc power-law relationship
between $T$ and $B_{r0}$ such that $T$ decreases as the star spins down as
$T\propto\Omega^\sigma$. In our $\alpha=1.1$ magnetized polytropic wind models,
a pronounced break in spin-down, as suggested by observations \citep[see, e.g., Fig.~2 in][]{vanSadersetal2016}, requires $\sigma \gta 1.5$. Direct extrapolation of such a steep power law to rapidly rotating young main-sequence solar-type stars would yield unphysically high coronal temperature. \cite{OFionnagainVidotto2018} have bypassed this difficulty by assuming a variation of coronal temperature with rotation rate described by a double power-law $T\propto \Omega^{1.2}$ for $\Omega<1.4\Omega_\odot$, switching to
$T\propto \Omega^{0.37}$ above that rotation period. Recovering Skumanich's Law using such a temperature dependency would then require the introduction of other ad hoc dependencies in other wind parameters, for example the wind base density.

In polytropic winds models, a class to which the WD model belongs, the energy input is set implicitly by both the assumed coronal base temperature and value for the polytropic index. We demonstrated in \S\ref{sec:Wenergetics} that close to the hydrostatic corona limit, the WD wind model exhibits a very tight correlation between mass loss rate and total power input into the wind, so that the ad hoc decreases in coronal temperature required to break the Skumanich $t^{-1/2}$ relationship in
\S\ref{sec:breakSku} implies a drop
in power input by over three orders of magnitude as compared to that estimated for the corona of the present-day sun.
Observational evidence for variations of chromospheric emission beyond the solar age \citep[e.g.,][and references therein]
{Lorenzo-Oliveiraetal2018} show no evidence for such a drastic drop, although care is warranted considering that the bulk of the observed emission presumably arises in compact, closed magnetic fields overlying active regions, as it does on the sun. To what extent this reflects closely (or not) the temperature of (and power input to) the quiet solar corona remains an open question; between minimum and maximum phases of the solar magnetic activity cycle, chromospheric and coronal emission varies by orders of magnitude, while the temperature of the quiet corona does not. This fact alone indicates that inferring stellar \emph{quiet} coronal temperatures from non-thermal emission measures is a risky proposition, at the very least.

All such observational and physical uncertainties notwithstanding, the modeling results presented in this paper indicate that even under the most favorable conditions, namely a very steep drop of coronal temperature with decreasing global coronal magnetic field and rotation rate, thermal wind collapse alone is unlikely to fully explain the observationally-inferred very rapid drop in angular momentum loss in solar-type stars slightly older than the sun. This implies that additional physical mechanisms, such as a shutdown or topological reconfiguration of the dynamo-generated large-scale coronal magnetic field, must also play a significant role.

%Is idea of coronal wind collapse as explanation of break in gyrochronology viable or not ? discuss

%Prochain article: relation dynamo $B$ vs $\Omega$ in the context of spin-down from the ZAMS.

\begin{acknowledgments}
We wish to thank an anonymous referee for useful comments and suggestions.
This research is supported by NSERC Discovery grant RGPIN-2024-04050 and a graduate fellowship from Banque Nationale. Both authors are members of the Centre de Recherche en Astrophysique du Qu\'ebec (CRAQ) and of the Institut Trottier de Recherche sur les Exoplan\`etes (IREx).
\end{acknowledgments}

\appendix

\section{Hydrostatic polytropic coronae}

We consider a spherically symmetric unmagnetized hydrostatic
solar corona under the polytropic approximation
\begin{eqnarray} \label{eq:polytropic}
{p\over p_0}=\left({\rho\over\rho_0}\right)^\alpha~.
\qquad 1\leq\alpha\leq 5/3
\end{eqnarray}
where a subscript ``0'' indicates a quantity evaluated at a reference heliocentric radius
$r_0$ corresponding to the base of the corona. Its radial structure
is described by the equation of hydrostatic equilibrium in a
$1/r^2$ external gravity field, which in view of Eq.~(\ref{eq:polytropic}) takes the form:
\begin{eqnarray} \label{eq:drhodr}
c_{s0}^2\left({\rho\over\rho_0}\right)^{\alpha-1}{{\rm d}\rho\over {\rm d}r}
=-\rho{GM_\odot\over r^2}~,
\end{eqnarray}
where $c_{s0}^2=\alpha p_0/\rho_0$ is the base polytropic sound speed.
This readily integrates to
\begin{eqnarray}
\label{eq:density}
{\rho(r)\over\rho_0}=\left[1-{(\alpha-1)GM_\odot\over r_0c_{s0}^2}
\left(1-{r_0\over r}\right)\right]^{1/(\alpha-1)},
\end{eqnarray}
from which the pressure profile $p(r)/p_0$ follows immediately
using Eq.~(\ref{eq:polytropic}). Both pressure and density vanish in
the asymptotic limit $r\to\infty$ provided
\begin{eqnarray}
\label{eq:rtop}
c_{s0}^2\leq (\alpha-1)GM_\odot/r_0~.
\end{eqnarray}
in which case the corona extends to a finite radius $r_{\rm top}/r_0=(1-r_0c_{s0}^2/(\alpha-1)GM_\odot)^{-1}$.
If (\ref{eq:rtop}) is not satisfied,
then both pressure and density asymptote to constant values, implying coronal expansion
\citep{Parker1958}. For a perfect gas of mean molecular weight $\mu$,
the equation of state allows to express
$c_{s0}$ in term of the base coronal temperature $T_0$ as $c_{s0}^2=\alpha k_BT_0/\mu m_p$ which,
when substituted into Eq.~(\ref{eq:rtop}), leads to the constraint
(\ref{eq:hydrostatic}) on the base temperature setting the hydrostatic limit.

\section{Numerical solution of the Weber-Davis wind equations}

We follow \cite{BelcherMacGregor1976} (\S {I}{I}) in reducing the calculation of a Weber-Davis 
wind solution to a six-dimensional
nonlinear root finding problem, from which the full wind solution can be reconstructed. 
Straightforward (but somewhat tedious) algebraic manipulations allow to combine
the governing equations of the WD model ---the $r$- and $\phi$-component of the momentum and induction equations, the continuity equation,
and the polytropic relationship
$p\propto\rho^\alpha$--- into either of the two following forms:
\begin{eqnarray}
\label{eq:WD16}
{\partial u_r\over\partial r}=\left({u_r\over r}\right)
{
(u_r^2-A_r^2)(2c_s^2+u_\phi^2-GM/r)+2u_ru_\phi A_rA_\phi
\over
(u_r^2-A_r^2)(u_r^2-c_s^2)-u_r^2A_\phi^2
}~,
\end{eqnarray}
\begin{eqnarray}
\label{eq:WD15}
{1\over 2}(u_r^2+u_\phi^2)-{GM\over r}
+{c_{s0}^2\over\alpha-1}\left({\rho\over\rho_0}\right)^{\alpha-1}
-{r\Omega A_rA_\phi\over u_r}=E~,
\end{eqnarray}
The first is essentially a rewriting of the $r$-component of the momentum
equation, while the second is a first integral akin to Bernoulli's equation for this problem, with the constant $E$ corresponding to the total energy per unit 
mass in a fluid element moving along a flow streamline. The denominator ($D$) of
Eq.~(\ref{eq:WD16}) vanishes when the radial flow speed becomes equal to
the slow or fast magnetosonic wave speeds.
Denoting the radial positions and radial flow speeds at which this takes place
as $(r_s,u_{rs})$ and $(r_f,u_{rf})$, respectively, regularity of the solution
is enforced by requiring that the numerator ($N$) also vanishes at these
positions; this yields four coupled non-linear algebraic equations:
\begin{eqnarray}
\label{eq:WD17}
N(r_f,u_{rf})=0~,\qquad
D(r_f,u_{rf})=0~,\qquad
N(r_s,u_{rs})=0~,\qquad
D(r_s,u_{rs})=0~,
\end{eqnarray}
Equation (\ref{eq:WD15}) is then invoked to further require that the energy
per unit mass at these points be the same as at the base of the flow, which provides two additional nonlinear algebraic equations:
\begin{eqnarray}
\label{eq:WD18}
E(r_f,u_{rf})=E(r_0,u_{r0})~,\qquad
E(r_s,u_{rs})=E(r_0,u_{r0})~.
\end{eqnarray}
Solution of these six coupled nonlinear algebraic equations yields the solution vector $[u_{r0},u_{\phi0}, r_s, u_{rs}, r_f, u_{rf}]$.
Any one such solution is defined by a 5-dimensional input vector $[\alpha, c_{s0}, \gamma, \zeta, \beta ]$, grouping
%
%\begin{equation}
%[\alpha, c_{s0}, \gamma, \zeta, \beta ]~,
%\end{equation}
%
%these being
the polytropic index $\alpha$;
the base sound speed $c_{s0}$, measuring coronal temperature;
the gravitational escape speed, measuring gravity;
the azimuthal speed of the rotating frame, measuring rotation rate; and
the Alfv\'en speed, measuring coronal magnetic field strength.
All are
evaluated at the base radius $r_0$, and the latter three
normalized to the base polytropic sound speed $c_{s0}$.

%
%\textcolor{blue}{Solution of the six coupled nonlinear algebraic equations for the six unknowns jointly define the solution vector $[u_{r0},u_{\phi0}, r_s, u_{rs}, r_f, u_{rf}]$,
%
%\begin{equation}
%[u_{r0},u_{\phi0}, r_s, u_{rs}, r_f, u_{rf}]~,
%\end{equation}
%
%where $u_{r0},u_{\phi 0}$ are the radial and azimuthal components of the base flow speed, $r_s,r_j$ are the positions of the slow and fast magnetosonic points, and
%$u_{rs},u_{rf}$ the corresponding radial flow speeds.}
%

Strongly nonlinear multidimensional root finding problems of this type are notoriously challenging from a numerical point of view \citep[see, e.g., the insightful discussion in \S 9.6 of][also \S 4.2 in \citealt{Charbonneau1995} for an alternate approach based on nonlinear minimization]{Pressetal1992}. Standard gradient-based schemes, such as Newton-Raphson, can be used; however, in the present case, the radius of convergence can be quite small, requiring a careful setting of the step length and a good initial guess to initiate a successful steepest descent to the sought-after solution.

In the context of spin-down calculations in which a large number of WD solutions need to be calculated to compute the angular momentum loss rate ${\dot J}$ along any given spindown sequence, we have opted
to pre-compute a grid of WD solutions in a 4-dimensional input parameter
space. For the work presented in this paper, this grid covers the (physical)
ranges
$\alpha \in [1.05, 1.15]$, $T\in  [0.81,4.50]\,$MK, $\Omega/\Omega_\odot \in [0.6,25]$, $B\in [1.67, 69.8]$ G,
%$B\in [.6,25] B_\odot$ [$1.67, 69.8$] G,
%The normalization here changes the parameters range. This is caused by the temperature adjustment of our dimensional parameters. [On ne parle pas d'adimensionalisation dnas l'article, donc je ferais sauter ce bout de phrase pour le moment]
and gravity fixed at the solar value.

Starting from a solar solution, the grid is filled by increasing sequentially one
of the four input parameters, and using the previous member of the sequence as the initial starting guess for the current member. %With small enough steps through
%parameter space, this continuation schemes functions well, except in some extreme corners of parameter space, e.g. high rotation rate and low magnetic field strengths, or low temperature and polytropic index close to unity, in which case finer steps are required.
%The solution grid is then resampled
%on a uniform 4-dimensional grid, and 
Interpolation is then used to extract a solution for any combination of input parameters.
This procedures introduces a small numerical error, but in practice for a tight enough
pre-computed grid it remains quite small and insignificant for mass and angular momentum loss computation. Significant interpolation errors are readily detected when reconstructing full radial wind profiles, as they
typically manifest themselves as spurious jumps across energetically allowed solution branches, resulting from failure to properly cross a critical points.

\bibliographystyle{aasjournal}
%\bibliography{Paulrefs.bib}{}

\end{document}